\documentclass[sort&compress,final,3p,times,twocolumn,fixfloat]{elsarticle}
\usepackage{comment}
\usepackage{graphicx,color}

\journal{Physics Letters B}

\def\ie{{\it i.e.}}
\def\eg{{\it e.g.}}

\newcommand{\ce}[1]{Eq.~(\ref{#1})}
\newcommand{\cf}[1]{{Fig.~\ref{#1}}}
\newcommand{\ct}[1]{{Table~\ref{#1}}}

\newcommand{\msbar} {\overline{\text{MS}}}

\def\alphaS{\alpha_s}

\usepackage{txfonts}
\usepackage{bm}
\usepackage{amsmath}
\usepackage{amssymb}
\usepackage{booktabs}
\usepackage{float}
\usepackage{comment}
\usepackage{lineno}
\usepackage{xspace} 
\usepackage{mathtools, cuted}

\usepackage{scalerel}
\usepackage{tikz}
\usetikzlibrary{svg.path}

\DeclareMathAlphabet{\pazocal}{OMS}{zplm}{m}{n}
\newcommand{\Q}{\pazocal{Q}}

\usepackage[normalem]{ulem}

\newcommand{\new}[1]{{\color{black}#1}}

\def\jpsi    {\mbox{$J/\psi$}}

\newcommand*\oline[1]{%
   \vbox{%
     \hrule height 0.5pt%
     \kern0.4ex%
     \hbox{%
       \kern-0.15em%
       \ifmmode#1\else\ensuremath{#1}\fi%
       \kern-0.15em%
     }%
   }%
}

 \usepackage{ifpdf}
 \ifpdf
 \usepackage[pdftex]{hyperref}
 \else
 \usepackage[hypertex]{hyperref}
 \fi

 \hypersetup{
   pdftitle={},%
   pdfauthor={},%
   pdfsubject={},%
   pdfkeywords={},%
   pdfstartview={},%
   bookmarksopen=true, breaklinks=true, debug=true, %
   colorlinks=true, linkcolor=red, citecolor=blue, urlcolor=blue
 }

\usepackage{float} %
\usepackage[caption=false]{subfig} %
\usepackage[sort&compress, capitalise]{cleveref}

\newcount\savefnused
\newcount\savefndone

\newcommand{\savefootnote}[2][\empty]%
{\ifx\empty#1\footnotemark\else\footnotemark[#1]\fi
 \global\advance\savefnused by 1
 \expandafter\xdef\csname savefnmark\the\savefnused\endcsname{\thefootnote}%
 \expandafter\xdef\csname savefntext\the\savefnused\endcsname{#2}%
}
\newcommand{\flushfootnote}{\loop\ifnum\savefndone<\savefnused
  \global\advance\savefndone by 1
  \footnotetext[\csname savefnmark\the\savefndone\endcsname]%
    {\csname savefntext\the\savefndone\endcsname}%
  \global\expandafter\let\csname savefnmark\the\savefndone\endcsname\relax
  \global\expandafter\let\csname savefntext\the\savefndone\endcsname\relax
\repeat}

\begin{document}

\title{Revisiting NLO QCD corrections to total inclusive $J/\psi$ and $\Upsilon$ photoproduction cross sections in lepton-proton collisions}

\date{03-11-2022}

\address[IJCLab]{Universit\'e Paris-Saclay, CNRS, IJCLab, 91405 Orsay, France}
\address[Chongqing]{Department of Physics, College of Basic Medical Sciences, Army Medical University, Chongqing 400038, China}
\address[Karlsruhe]{Institute for Theoretical Particle Physics, KIT, 76128 Karlsruhe, Germany}
\address[LPTHE]{Laboratoire de Physique Th\'eorique et Hautes Energies, UMR 7589,
Sorbonne Universit\'e et CNRS, 4 place Jussieu, 75252 Paris Cedex 05, France}
\address[UCD]{School of Physics, University College Dublin, Dublin 4, Ireland}

\author[IJCLab]{Alice Colpani Serri}

\author[Chongqing]{Yu Feng}

\author[IJCLab]{Carlo Flore%
}
\ead{carlo.flore@ijclab.in2p3.fr}

\author[IJCLab]{Jean-Philippe Lansberg%
}
\ead{jean-philippe.lansberg@in2p3.fr}

\author[IJCLab,Karlsruhe]{Melih A. Ozcelik%
}
\ead{melih.oezcelik@kit.edu}

\author[LPTHE]{Hua-Sheng Shao%
}
\ead{huasheng.shao@lpthe.jussieu.fr}

\author[IJCLab,UCD]{Yelyzaveta Yedelkina%
}
\ead{yelyzaveta.yedelkina@universite-paris-saclay.fr}

\begin{abstract}
We revisit inclusive $J/\psi$ and {$\Upsilon$} photoproduction at lepton-hadron colliders, namely in the limit when the exchanged photon is quasi real. Our computation includes the next-to-leading-order (NLO) $\alpha_s$ corrections {to the leading-order contributions in $v$}. Similarly to the case of NLO charmonium-hadroproduction processes, the resulting cross sections obtained in the $\overline{\text{MS}}$ factorisation scheme are sometimes found to be negative. We show that the scale-fixing criteria which we derived in a previous study of $\eta_c$ production successfully solves this problem {from the EicC all the way up to the FCC-eh energies}. \new{We then elaborate on how to study a scale uncertainty akin to that derived by scale variations when one fixes a scale.} 
In turn, we investigate where both $J/\psi$ and $\Upsilon$ photoproduction could be used to improve our knowledge of gluon {content of the proton} at scales as low as a couple of GeV. 
\end{abstract}

\begin{keyword}
$J/\psi$, $\Upsilon$, $ep$ reactions, photoproduction, HERA, EIC, EicC, AMBER-COMPASS++, LHeC, FCC-eh
\end{keyword}
\maketitle

\section{\label{sec:intro}Introduction}
Inclusive\footnote{also referred to as inelastic} production of quarkonia in hadron-hadron and lepton-hadron collisions is a potential rich source of information on the hadron structure. As such, it has been thoroughly studied both experimentally and theoretically (see ~\cite{Lansberg:2019adr,Andronic:2015wma,Brambilla:2010cs,Lansberg:2006dh, Brambilla:2004wf,Kramer:2001hh} for reviews). Yet, the mechanisms underlying their inclusive production {are} still not an object of consensus within the community. This in turn does not {encourage} one to employ cross-section measurements to extract information on the gluon structure of the proton.

In a recent study~\cite{Flore:2020jau,Flore:2021rlc}, we have however shown that the large-$P_T$ inclusive $\jpsi$ photoproduction data can be  accounted for by the Colour-Singlet Model {(CSM)}~\cite{Chang:1979nn,Berger:1980ni,Baier:1983va}, \ie~the leading-$v$ contribution of {Non-Relativistic QCD (NRQCD)~\cite{Bodwin:1994jh}}. In the photoproduction limit, a quasi on-shell photon hits and breaks a proton to produce the $\jpsi$ {usually along}  with at least a recoiling hard parton. This limit has been studied in detail at HERA~\cite{Aid:1996dn,Breitweg:1997we,Chekanov:2002at,Adloff:2002ex,Chekanov:2009ad,Aaron:2010gz,Abramowicz:2012dh} to decipher the quarkonium-production mechanisms and then to probe the gluon content of the proton (see \eg~\cite{Jung:1992uj}). As expected, photoproduction indeed seems to be more easily understandable than hadroproduction~\cite{Lansberg:2019adr}\footnote{At high energies, the hadronic content of the photon can be ``resolved'' during the collisions. Resolved-photon -- proton collisions are very similar to those for hadroproduction and are interesting on their own. We will however disregard them in the present discussion and they can be avoided by a simple kinematical cut on low elasticity values, $z$. Along the same lines, exclusive or diffractive contributions can be also avoided by cutting $z$ close to unity.}. It is also believed that quarkonium production in lepton-proton collisions could be used to measure Transverse Momentum Dependent gluon distributions (see \eg~\cite{Chapon:2020heu,Boer:2020bbd,Kishore:2019fzb,DAlesio:2019qpk,Bacchetta:2018ivt}).

Owing to the presence of an electromagnetic coupling, photoproduction cross sections are smaller than hadroproduction ones which calls for large luminosities to obtain large enough quarkonium data sets. As such, the $P_T$ reach of $J/\psi$ HERA data is limited to barely 10 GeV, there is quasi no data on $\psi'$ and none on the $\Upsilon$.

In the present analysis, we focus on the $P_T$-integrated {yield} which was surprisingly seldom studied at HERA. Indeed most of the inclusive data set have been selected with the minimal $P_T$ of 1 GeV. This {cut} however  {introduces} a strong sensitivity on the $P_T$ spectrum of the cross section in a region where it is not necessarily well controlled. {By itself}, {such a yield} is not directly connected to the total number of $J/\psi$ produced for which we believe the theory predictions to be more robust. 
{One} reason for such a cut is probably the difficulty to obtain numerically stable NLO results when they appeared~\cite{Kramer:1994zi,Kramer:1995nb}\footnote{In particular, we note the reference [48] of~\cite{H1:1996kyo}.}. In fact, as we will discuss, these were probably due to the appearance of large negative NLO contributions which we will address here along the same lines as our recent study on $\eta_c$ production~\cite{Lansberg:2020ejc}.

Moreover, such $P_T$-integrated cross sections will be easily measurable at high energies with a very good accuracy at the planned US Electron-Ion Collider (EIC)~\cite{Accardi:2012qut}, but also at future facilities such as the LHeC~\cite{LHeCStudyGroup:2012zhm} or FCC-eh~\cite{FCC:2018byv}, thus in a region where the gluon PDFs are not well constrained. Measurements at lower energies at AMBER-COMPASS++~\cite{Adams:2018pwt} and the EicC~\cite{Anderle:2021wcy} would then rather probe the valence %
region, which could happen to be equally interesting.

In our study, like in the previous one~\cite{Flore:2020jau}, we will focus on the aforementioned CSM~\cite{Berger:1980ni}, corresponding to the leading-$v$ contribution {in} NRQCD whose NLO QCD corrections are in principle known since the mid nineties~\cite{Kramer:1994zi,Kramer:1995nb}. %
As we revisited in~\cite{Flore:2020jau},  the impact of QCD corrections to $J/\psi$ inclusive photoproduction steadily grows when $P_T$ increases. This can be traced back to the more favourable $P_T$ scaling of specific real-emission contributions. The same has been observed in several quarkonium-hadroproduction processes~\cite{Artoisenet:2008fc,Lansberg:2008gk,Lansberg:2009db,Gong:2012ah,Lansberg:2013qka,Lansberg:2014swa,Lansberg:2017ozx,Shao:2018adj}. On the contrary, one expects, at low $P_T$, a more subtle interplay between the contribution of these real emissions near the collinear region and the loop corrections. {This} has for a long time {been} {understudied}. {We thus aim at discussing it} here in detail.

The structure of our Letter is as follows. Section~\ref{sec:methods} outlines our methodology to compute vector-quarkonium inclusive photoproduction cross sections at NLO accuracy including a discussion of the reason for negative NLO cross sections\new{, our proposed solution and a discussion on how to account for scale uncertainties in a computation when one scale is fixed}. Section~\ref{sec:results} gathers our prediction for future measurements at lepton-hadron colliders {along with a discussion of the corresponding theoretical uncertainties}. Section~\ref{sec:conclusions} gathers our conclusions.

\section{\label{sec:methods}$J/\psi$ and $\Upsilon$ photoproduction up to one-loop accuracy}

\subsection{Elements of kinematics}
We will consider the process $\gamma(P_\gamma)+p(P_p) \to \Q(P_\Q) + X$ where the photon $\gamma(P_\gamma)$ is emitted by an electron $e(P_e)$.
Let us then define $s_{ep}=(P_e + P_p)^2\approx 4 E_e E_p + m_p^2$ %
($E_{e(p)}$ is the electron (proton) beam energy, $m_p$ is the proton mass) and $s_{\gamma p}=W_{\gamma p}^2 = (P_\gamma + P_p)^2$. We can then introduce $x_\gamma$ as $P_\gamma= x_\gamma P_e$ such as $s_{\gamma p}=x_\gamma s_{ep}$. As announced, in the present study, $P^2_\gamma\simeq 0$.

Diffractive contributions are suppressed for increasing $P_T$ and away from the exclusive limit, \ie\ when the quarkonium carries {nearly} all the photon momentum. A cut on $P_{\Q T}$ is usually sufficient to get rid of them, which we do not wish to apply here. One can however cut on a variable called elasticity, defined as $z = \tfrac{P_{\Q}\cdot P_p}{P_\gamma \cdot P_p}$. $z$ indeed corresponds to the fraction of the photon energy taken by the quarkonium in the proton rest frame, with the proton momentum defining the z axis.
It can be rewritten as $z=\tfrac{2\,E_p \,m_T}{W^2_{\gamma p}\,e^{y}}$ in terms of the quarkonium rapidity $y$ (with $y$ and $E_p$ being defined in the same frame) and  the quarkonium transverse mass, $m_T = \sqrt{{M}_\Q^2 + P_{\Q T}^2}$ {with $M_\Q$ being the quarkonium mass}. Such {diffractive} contributions are known to lie at $z\to 1$~\cite{Aid:1996dn}. {At low $z$, resolved-photon contributions can appear as important where only a small fraction of the photon energy is involved in the quarkonium production.  At HERA, they had a limited impact~\cite{Aid:1996dn,Kramer:2001hh}. At lower energies, like at the EIC, their impact should be further reduced. On the contrary, at the LHeC or FCC-eh, their impact might be sizable even at moderate $z$. Nonetheless, their modelling requires a good control of the contributions from $gg$ and $gq$ channels.
{However, our understanding of the very same channels in inclusive quarkonium hadroproduction, especially at low $P_T$~\cite{Lansberg:2019adr}, is clearly limited.} When comes the time for the building of these future lepton-hadron colliders, it will be needed to re-evaluate the impact of the resolved-photon contributions at low $z$ and high energies. For the time being, we simply note that imposing a lower bound on $z$ would not alter our conclusions at all, precisely because it does not correspond to the low-$x$ region in the proton.}

\subsection{The Colour-Singlet Model}

As {mentioned earlier}, there is no agreement on which mechanism is dominant in quarkonium production. The most popular approaches are: the Colour-Singlet Model (CSM)~\cite{Chang:1979nn,Berger:1980ni,Baier:1983va}, the Colour-Evaporation Model (CEM)~\cite{Fritzsch:1977ay,Halzen:1977rs} and the Non-Relativistic QCD (NRQCD)~\cite{Bodwin:1994jh}, whose leading-$v$ contribution is the CSM {for $S$-wave quarkonia}. These mechanisms  mainly differ in the way they describe hadronisation. 
The factorisation approach of the CSM is based on considering only the leading Fock states of NRQCD. In the CSM, there is no gluon emission during the hadronisation process, and, consequently, the quantum state $Q\overline{Q} $ does not evolve during the binding. Thus, spin and colour remain unchanged.

The matrix element $ {\cal M}$ to produce a vector state $\Q$ + $\{k\}$, where $\{k\}$ is a set of final state particles, from the scattering of the partons $ab$ in CSM is: 
\begin{equation}
\begin{aligned}\label{eq:amplitude}
 {\cal M}(ab \rightarrow \Q+\{k\})=\sum_{s_1,s_2,i,i'} \frac{N(\lambda | s_1,s_2)}{ \sqrt{m_Q}} 
\frac{\delta^{ii'}}{\sqrt{N_c}} 
\frac{R(0)}{\sqrt{4 \pi}} \times \\
\times  {\cal M}(ab \to Q^{s_1}_i \bar Q^{s_2}_{i'}(\mathbf{p}=\mathbf{0})  + \{k\}),
\end{aligned}
\end{equation}
where {$N(\lambda| s_1,s_2)$ (resp. $\delta^{ii'}/\sqrt{N_c}$) is the projector onto a vector (resp. CS) state}  and $ {\cal M}(ab \to Q \bar Q+\{k\})$ is the amplitude to create the corresponding heavy-quark pair. When one then sums  over the heavy-quark spins, one obtains usual traces which can be evaluated without any specific troubles. In fact, such a computation can be automated at tree level as done by {\sc HELAC-Onia}~\cite{Shao:2012iz, Shao:2015vga}. In the present case of inclusive photoproduction of a vector $\Q$, there is a single partonic process at Born order, $\alpha \alphaS^2$, namely $\gamma g \to \Q g$ (see~\cf{lo}). One could also consider $\gamma Q \to \Q Q$ at the same order, but we have shown it to be small at low $P_T$~\cite{Flore:2020jau}.

The value of $R(0)$, the $\Q$ radial wave function at the origin in the configuration space, can be {in principle} extracted from the leptonic decay width computed likewise in the CSM. {The latter is known up to NLO~\cite{Barbieri:1975ki} since the mid 70's,  up to NNLO since the late 90's~\cite{Czarnecki:1997vz,Beneke:1997jm} and up to N$^3$LO since 2014~\cite{Marquard:2014pea}.
However, as discussed in~\ref{sec:LDME}, the short-distance amplitude receives very large QCD radiative corrections which translate into significant renormalisation and NRQCD-factorisation scale uncertainties. These essentially preclude drawing any quantitative constraints on $|R(0)|^2$ from the leptonic decays width.}

In principle, we should thus associate to it a specific theoretical uncertainty which is however supposed to only affect the normalisation of the cross sections. In what follows, we will employ {a similar} value as Kr\"amer~\cite{Kramer:1995nb}, {1.25}~GeV$^3$ for the $J/\psi$ and 7.5 GeV$^3$ for the $\Upsilon(1S)$.
As for the masses, we will use $m_c=1.5$~GeV and $m_b=4.75$~GeV. {Let us recall that within NRQCD $M_\Q=2m_Q$.}

\begin{figure}[hbt!]
\centering  \captionsetup[subfloat]{captionskip=-0.05cm}
\subfloat[]{\includegraphics[scale=.23]{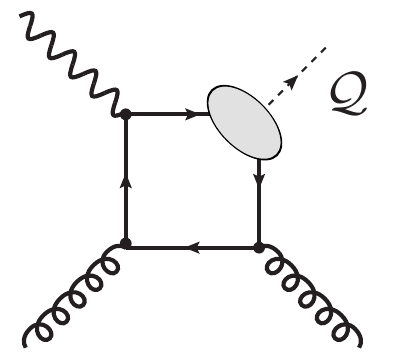}\label{lo}}
\subfloat[]{\includegraphics[scale=.23]{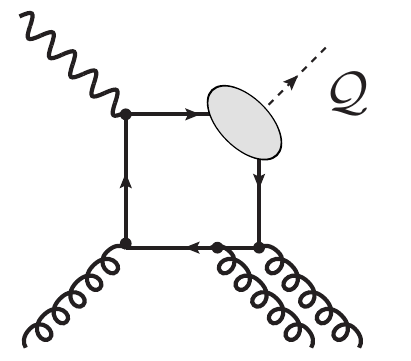}\label{q_gl}}
\subfloat[]{\includegraphics[scale=.23]{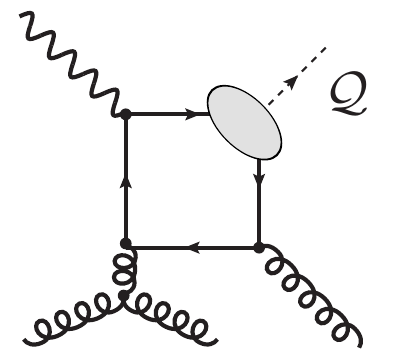}\label{init_gl_gl}}
\subfloat[]{\includegraphics[scale=.23]{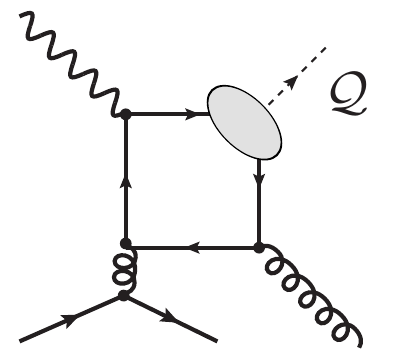}\label{init_gl_qq}}\\\vspace*{-0.25cm}

\subfloat[]{\includegraphics[scale=.23]{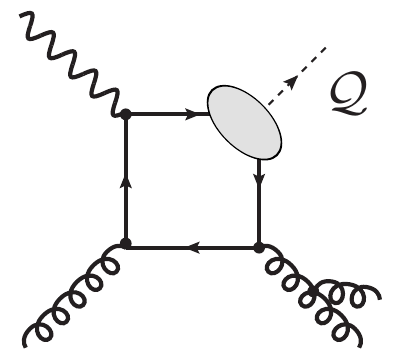}\label{fin_gl_gl}}
\subfloat[]{\includegraphics[scale=.23]{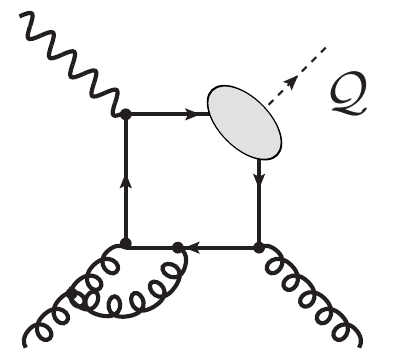}\label{vtx_corr}}
\subfloat[]{\includegraphics[scale=.23]{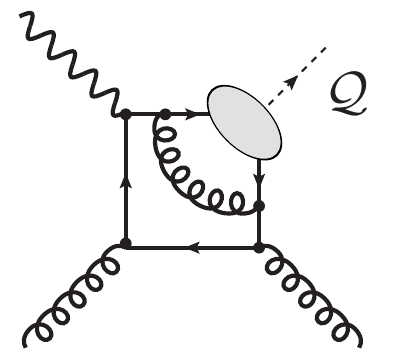}\label{box}}
\subfloat[]{\includegraphics[scale=.23]{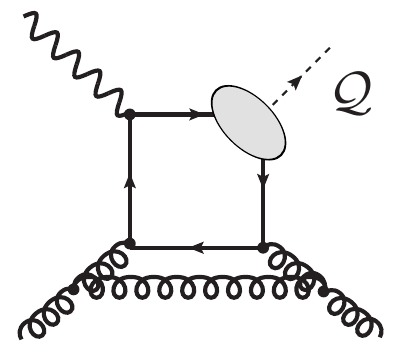}\label{loop}}\vspace*{-0.15cm}
\caption{Representative Feynman diagrams for inelastic $\Q$ %
photoproduction contributing via CS channels at orders $\alpha \alpha_s^2$ (a), $\alpha \alpha_s^3$ (b,c,d,e,f,g,h). The quark and anti-quark attached to the ellipsis are taken as on-shell and their relative velocity $v$ is set to zero.\vspace*{-0.35cm}}
\label{diagrams-CSM-photoproduction}
\end{figure}

{\cf{lo} displays one of the six Feynman diagrams for inelastic $\Q$ photoproduction at LO ($\alpha \alphaS^2$).
At this order, only photon-gluon fusion contributes. 

The hadronic cross section is readily obtained by folding the partonic cross section, $d\hat{\sigma}_{\gamma g}^{(0)}$,
with the corresponding PDFs and, if relevant, summing over the parton species.
 Generically, one has:
\begin{equation}
   d\sigma_{\gamma p}(s_{\gamma p},m_Q^2)=\sum_{i=g,q,\bar{q}}\int\!\! dx f_{i}(x,\mu_F) d\hat{\sigma}_{\gamma i}.
\label{collinear_f}
\end{equation}
where $\mu_F$ is the factorisation scale {and} $f_{i}(x, \mu_F)$ is the PDF that gives the probability that a parton $i$ carries a momentum fraction $x$ of the parent proton. At LO, $d\hat{\sigma}_{\gamma i}$ {identifies to} $d\hat{\sigma}_{\gamma g}^{(0)}$.%

\subsection{The NLO corrections and their divergences}

At order $\alpha \alphaS^3$, two categories of new contributions arise. Those from the real emissions represented by \cf{q_gl},\ref{init_gl_gl}, \ref{init_gl_qq}, \ref{fin_gl_gl} and from virtual emissions (or loops) represented by \cf{vtx_corr}, \ref{box}, \ref{loop}. Specific topologies of the former category benefit from $P_T/M_\Q$ enhancement factors which make them leading at large $P_T$. This for instance justifies to employ the NLO$^\star$ approximation~\cite{Flore:2020jau}. When $P_T$ is integrated over, a priori all these contributions should be accounted for. Such a computation was first carried out by Kr\"amer~\cite{Kramer:1994zi} in the mid 1990's. We will briefly outline now what it amounts to.

As usual, one critical feature of such NLO computations in collinear factorisation~\cite{CTEQ:1993hwr} is the appearance of various types of singularities. Of particular relevance for our discussion are the collinear divergences  from initial-state emissions which arise when one cannot distinguish two massless particles, with an angle between them close to zero. 
These remain because the initial states are fixed by the kinematics of collinear factorisation
and, consequently, are not fully integrated over. {These singularities are absorbed inside the $\msbar$-renormalised PDFs via the process-independent Altarelli-Parisi Counter Terms (AP-CT).} 
These AP-CT introduce a $\mu_F$ dependence in the partonic cross section, which would, in an all-order computation, cancel that introduced by the PDF scale evolution governed by the DGLAP equation.

\subsection{The cross section in terms of scaling functions}

For our analysis, we have found useful to employ the NLO cross-section decomposition in terms of scaling functions derived by Kr\"amer~\cite{Kramer:1995nb}.  Using FDC~\cite{Wang:2004du,Gong:2012ah}, we have reproduced his results (scaling functions as well as hadronic cross sections) and, with the appropriate parameter choices and kinematical cuts, those of~\cite{Butenschoen:2009zy}.

Indeed, the advantage of considering $P_T$- and $z$-integrated cross sections is that the hadronic photoproduction cross sections can be recast in terms of a simple convolution of the PDF
and scaling functions of a single scaling variable\footnote{In what follows, we will show them as a function of $\hat{s}$ and $m_Q$ but they can equally be written as a function of $\eta=\hat{s}/4m_Q^2-1$.}. This allows one to outline the structure of the result  to better understand some specific behaviour (like the scale dependencies discussed in the previous section, hence the importance of negative contributions to the cross section) of the NLO yield.
This formulation is also useful because it allows one to {economically vary} parameters like the c.m.~energy, the heavy-quark mass, the renormalisation and factorisation scales.

Along the lines of Kr\"amer~\cite{Kramer:1995nb}, we express the partonic cross section as\footnote{The scaling functions were derived in the $\msbar$ factorisation scheme. {Here $\overline{c}^{(1)}$ and $c^{(1)}$ correspond to $-\overline{c}^{(1)}$ and $c^{(1)}+\ln{4}\overline{c}^{(1)}$ defined by Kr\"amer~\cite{Kramer:1995nb}.}}: %
\begin{equation}
\begin{aligned}
&\hat\sigma_{\gamma i}(\hat{s},m_Q^2,\mu_R,\mu_F) =
\frac{\alpha\alpha_s^2(\mu_R) e^2_Q}{m_Q^2}\,\frac{|R(0)|^2}{4\pi m_Q^3} \times \\
& \times  \Bigg[ c_{\gamma i}^{(0)}(\hat{s},m_Q^2) + 4\pi\alpha_s(\mu_R)   \Bigg\{c_{\gamma i}^{(1)}(\hat{s},m_Q^2)
+\overline{c}^{(1)}_{\gamma i}(\hat{s},m_Q^2){\ln\frac{M_{\Q}^2}{\mu_F^2} }\\&
+\frac{\beta_0(n_{lf})}{8\pi^2}c_{\gamma i}^{(0)}(\hat{s},m_Q^2)\ln\frac{\mu_R^2}{\mu_F^2}\Bigg\}\Bigg],
\end{aligned}\label{eq:sigma_scaling_functions}
\end{equation}
where $i=g,q,\overline{q}$, $\beta_0(n_{lf})=(11N_c-2n_{lf})/3$, with $n_{lf}$ the number of active (light) flavours. The scaling functions are shown on \cf{fig:sc_func}.
$c_{\gamma g}^{(0)}$ {arises from} the $\alpha \alphaS^2$ (LO) $\gamma g$ contributions, while
$c_{\gamma g}^{(1)}$ and $\overline{c}_{\gamma g}^{(1)}$ from the $\alpha \alphaS^3$ (NLO) $\gamma g$ contributions and
$c_{\gamma q}^{(1)}$ and $\overline{c}_{\gamma q}^{(1)}$ from the $\alpha \alphaS^3$ (NLO) $\gamma q$ contributions. {
$c_{\gamma g}^{(1)}$ encapsulates contributions\footnote{To be exact, the corresponding term should in principle exhibit a $n_f$ dependence from $\gamma g \to \Q q \bar q$. The difference between the case $J/\psi$ and $\Upsilon(1S)$ would be from $\gamma g \to \Q c \bar c$ with $m_c=0$ which can safely be neglected.} from both real- and virtual emissions. If it had contained only virtual contributions, it would scale like $c_{\gamma g}^{(0)}$ and eventually vanish at large $\hat{s}$. This {implies} that the asymptotic value of $c_{\gamma g}^{(1)}$ entirely comes from the real emissions.  
$\overline{c}_{\gamma g}^{(1)}$ only includes real emissions and comes along with an explicit $\mu_F$ dependence from the AP-CT.} {The last term, whose form is generic, comes from the renormalisation procedure.} 
The hadronic cross section is then obtained according to \ce{collinear_f}.

\begin{figure}[hbt!]
    \centering
\includegraphics[width=0.85\columnwidth]{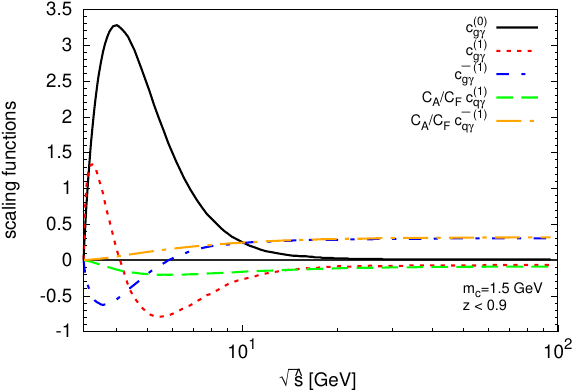}\label{fig:sc_func_q}\vspace*{-0.4cm}
\caption{Scaling functions as function of $\sqrt{\hat{s}}${, where $C_A=3,~C_F=4/3$.}\vspace*{-0.25cm} 
}
    \label{fig:sc_func}
\end{figure}

{Already at this stage, one can note then that, at large $\hat{s}$, the NLO cross section will be proportional to {$\ln({M_{\Q}^2}/{\mu_F^2})$} and a process-dependent coefficient, $\overline{c}^{(1)}_{\gamma i}(\hat{s}\to \infty,m_Q^2)$, which only comes on the real-emission contributions, like for $\eta_Q$ hadroproduction~\cite{Lansberg:2020ejc}.} %

\cf{fig:my_label} shows the $\sqrt{s_{\gamma p}}$-dependence of $\sigma_{\gamma p}$ for $J/\psi$ photoproduction integrated over  $z<0.9$ and $P_T$, for different choices of $\mu_R$ and $\mu_F$ among $M_{J/\psi} \times (0.5,1,2)$,
using the CT18NLO PDF set~\cite{Hou:2019efy} and with {a $20\%$ feed-down contribution from $\psi'$ decay (like in~\cite{Flore:2020jau}). We expect the $b$ feed down on the $P_T$-integrated yields to be on the order of 5\% and we do not include it as it can be experimentally removed.} 
For $\Upsilon (1S)$ photoproduction, we have estimated\footnote{These contributions were estimated using $|R_{\Upsilon(2S)}(0)|^2 = 5.0$~GeV$^3$  and $|R_{\Upsilon(3S)}(0)|^2 = 3.4$~GeV$^3$ and the corresponding measured branching fractions to $\Upsilon (1S)$~\cite{ParticleDataGroup:2020ssz}.} the feed-down contributions from $\Upsilon(2S)$ to be 12.5$\%$ and from $\Upsilon (3S)$ to be  2.2$\%$.

\begin{figure}[hbt!]
    \centering
\includegraphics[width=0.85\columnwidth]{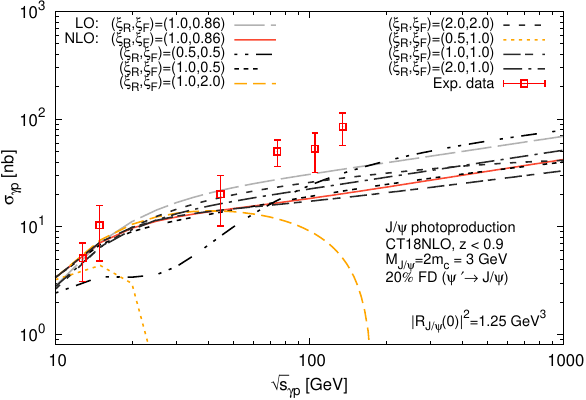}\vspace*{-0.4cm}
\caption{ LO and NLO $\sigma_{\gamma p}$ as a function of $\sqrt{s_{\gamma p}}$ for $J/\psi$ photoproduction for different scale choices (with the notation $\xi_{R,F} \equiv \mu_{R,F}/M_{J/\psi}$) compared with experimental data: H1~\cite{Aid:1996dn}, FTPS~\cite{Denby:1983az}, {NA14}~\cite{NA14:1986mdd}.\vspace*{-0.25cm} 
}
    \label{fig:my_label}
\end{figure}

The {long dashed} grey curve is the LO cross section for $\mu_R=M_\Q$ and $\mu_F=0.86~M_\Q$. We have checked that it remains positive for any $\mu_R$ and $\mu_F$ scale choice which is expected provided that the PDFs are positive. It happens to reasonably account for the available experimental values if one notes that the theoretical uncertainties from the scales, the mass and {$R(0)$} (not shown) are significant. 
All other curves represent the NLO cross section for different scale choices. {In two cases, the NLO cross section becomes negative as $s_{\gamma p}$ increases.}  {Anticipating the results of the next section, the same behaviour is obtained with other well-known PDF sets (MSHT20~\cite{Bailey:2020ooq}, NNPDF31~\cite{NNPDF:2017mvq}).} {As for $\Upsilon(1S)$, the cross section remains positive in the considered energy range for any realistic scale choice, like for $\eta_b$ hadroproduction up to 100~TeV~\cite{Lansberg:2020ejc}. Let us now discuss the origin of such an unphysical behaviour for $J/\psi$ photoproduction and propose a solution to it.}

\subsection{A new scale prescription to cure the unphysical behaviour of the NLO quarkonium photoproduction cross section}
\label{sec:muf-local-unc}

From the above discussion, there can only be two sources of negative partonic cross sections: the loop amplitude via interference with the Born amplitude and the real emissions via the subtraction of the IR poles from the initial-emission collinear singularities. As we will argue now, the latter subtraction is the source of {the negative cross section which we have just uncovered.}
As it was mentioned before, such divergences are removed by subtraction into the PDFs via AP-CT and the high-energy limit of the resulting partonic cross section takes the form:
\begin{equation}
\begin{aligned}
    \lim_{\hat{s}\rightarrow \infty}\hat{\sigma}_{\gamma i}^{\rm NLO}\propto\left({\log\frac{M_{\Q}^2}{\mu_F^2} +A_{\gamma i}} \right), A_{\gamma g}=A_{\gamma q},
\end{aligned}\label{eq:high-s_limit}
\end{equation} %
where $A_{\gamma i}={c_{\gamma i}^{(1)}(\hat{s}\to\infty,m_{Q}^2)}/{\overline{c}^{(1)}_{\gamma i}(\hat{s}\to\infty,m_{Q}^2)}$  are the coefficients of the finite term of NLO cross section in the high-energy limit. As can be seen from \cf{fig:sc_func}, $A_{\gamma i}$ is negative {for $z<0.9$, i.e. $-0.29$. It is also clear from \cf{fig:sc_func} that $A_{\gamma g}=A_{\gamma q}$.} %

Unless $\mu_F$ is sufficiently smaller than {$M_{\Q}$} in order to compensate $A_{\gamma i}$, $\lim\limits_{\hat{s}\rightarrow \infty}\hat{\sigma}_{\gamma i}^{\rm NLO}$ is negative,
like for $\eta_Q$~\cite{Lansberg:2020ejc} and it is another clear case
of oversubtraction by the AP-CT. Indeed, in this limit, the virtual contributions are suppressed; only the real emissions contribute via their square. As such, they can only yield positive partonic cross sections {\it before} the subtraction of the initial-state collinear divergences. Since, {\it after} their subtraction, the partonic cross section is negative, it has to come from the AP-CT. This is what we refer to as oversubtraction by the AP-CT.

In principle, the negative term from the AP-CT should be compensated by the evolution of the PDFs according to the DGLAP equation.
Yet, for the $\mu_F$ values on the order of the natural scale of these processes, the PDFs are not evolved much and can sometimes be so flat for some PDF parametrisations that the large $\hat s$ region still significantly contributes. This results in negative values of the hadronic cross section. Indeed, $A_{\gamma g}$ and $A_{\gamma q}$ are process-dependent, while the DGLAP equations are process-independent, which necessarily makes the compensation imperfect. Going to NNLO and even higher, this should naturally improve. If one has only NLO computations, this is however greatly problematic. A solution to this problem is~\cite{Lansberg:2020ejc} to force the partonic cross section to vanish in this limit, whose contribution should in principle be damped down by the PDFs.

According to this prescription, one needs to choose $\mu_F$ such  that $\lim\limits_{\hat{s}\rightarrow \infty}\hat{\sigma}_{\gamma i}^{\rm NLO}=0$. It happens to be possible since $A_{\gamma g} = A_{\gamma q}$. This amounts to consider that all the QCD corrections are in the PDFs~\cite{Lansberg:2020ejc}. From \ce{eq:high-s_limit}, we have:
\begin{equation}
\mu_F=\hat\mu_F= {M_{\Q}e^{A_{\gamma i}/2}=M_{\Q} {\rm exp}\left(\frac{c_{\gamma i}^{(1)}({\hat{s}\to \infty},m_{Q}^2)}{2 \overline{c}^{(1)}_{\gamma i}({\hat{s}\to \infty},m_{Q}^2)}\right)}. 
\end{equation}

Using the scaling function of \cf{fig:sc_func} when one fully integrates over $P_T$ and over $z<0.9$, one gets $\hat\mu_{F} = {0.86} M_{\Q}$. 
From now on, all our NLO results will be shown with this value of the factorisation scale. 

\begin{figure}[hbt!]
\includegraphics[width=0.85\columnwidth]{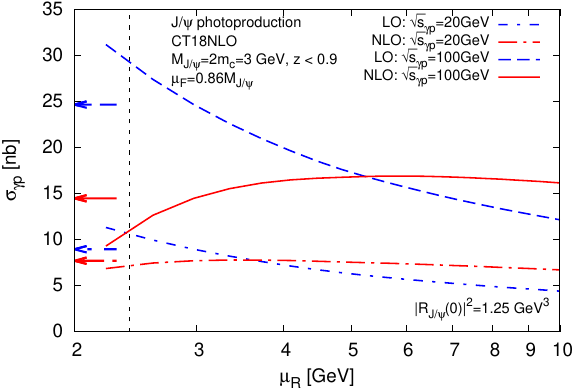}\vspace*{-0.4cm} 
\caption{$\mu_R$ dependence of $\sigma_{\gamma p}$ at LO and NLO for 2 values of $\sqrt{s_{\gamma p}}$, where the arrows point at the values of $\sigma_{\gamma p}$ for $\mu_R=3$ GeV. {The vertical dashed line delimitates the $\mu_R$ region which we use to compute the cross section (see text).\vspace*{-0.25cm} }
}
    \label{fig:mu_R/M}
\end{figure}

On \cf{fig:mu_R/M}, one can see the LO (in blue) and NLO (in red) $\mu_R$ dependence of $\sigma_{\gamma p}$ for $J/\psi$ photoproduction, still using CT18NLO and integrated over $P_T$ and over $z<0.9$, at two values of $\sqrt{s_{\gamma p}}=20$~GeV ({short and long} dash-dotted lines) and $\sqrt{s_{\gamma p}}=100$~GeV (solid {and dashed} lines).
In both cases, the $\mu_R$ sensitivity is drastically reduced at NLO. However, one notes that at the higher energy, for $\mu_R \sim M_{J/\psi}$, $\sigma^{\rm NLO}_{\gamma p}$  is twice smaller than {$\sigma^{\rm LO}_{\gamma p}$ (see the arrows by the $y$ axis). This is due to a large negative contribution from the loops (see the negative dip in the $c_{\gamma g}^{(1)}$ in \cf{fig:sc_func}).}
Since the LO and NLO cross section are however similar for $\mu_R \sim 2 M_{J/\psi}$, the question of the natural scale of the process naturally arises. In fact, as the Born process is $\gamma g \to \Q g$, it appears reasonable to consider $\sqrt{\hat{s}}$ rather than $M_\Q$. A quick LO computation for the $J/\psi$ case shows that $\sqrt{\langle \hat{s}\rangle}$ ranges from 4~GeV at low hadronic energies up to even 10~GeV at high hadronic energies.
\new{In what follows, we thus consider $\mu_R$ within the range $[2.5:10]$~GeV for $J/\psi$ and, for $\Upsilon(1S)$, $\mu_R \in [8:32]$~GeV (with $\mu_R = 5 (16)$ GeV being the center of this range in the two cases) at both LO and NLO and for $\mu_F$ at LO. The procedure for $\mu_F$ at NLO is discussed in the next section.}

\subsection{Scale fixing and theoretical uncertainties}
Just as we have discussed above, theoretical uncertainties of perturbative computations including radiative corrections arise from the appearances of unphysical scales, usually $\mu_R$ and $\mu_F$. In principle, predictions of physical observables, \eg\ a cross section, should not depend on them. This would only be so if all radiative corrections could be accounted for. At NLO, it is far from being the case and the scale dependences can be strong, so strong that some results are sometimes unphysical like in the process under discussion for some (reasonable) $\mu_F$ values. 

In general, the scale dependence is however mild (like for $\mu_R$ here) and evaluating observables at natural values of the scales, \ie\ those entering the kinematics of the process, yield to good predictions, which usually improve at NNLO and so forth. Since the scale dependences are meant to disappear in all-order computations, one can revert the argument and consider that the scale dependences give us some information about the impact of higher orders. This is why one varies the scale, like we have discussed at the end of the previous section, hoping to seize up some theoretical uncertainties from Missing-Higher Orders (MHO) in the jargon. This is however not done without ambiguity, to say the least, as the resulting uncertainties fully depend on the range of scale variation, conventionally chosen to be a multiplicative factor 2 for historical but unclear reasons\footnote{This practice might come from \cite{Altarelli:1988qr}.}. Whatever the variation should be, one then faces an apparent impossibility with our scale-fixing prescription: how to vary a scale which is fixed? 

To address this issue, it is necessary to go back to the very motivation of scale-fixing criteria. Like for our scale prescription, the other prescriptions (PMS~\cite{Stevenson:1981vj}, BLM-PMC~\cite{Brodsky:1982gc,Mojaza:2012mf,Wu:2013ei}, FAC~\cite{Grunberg:1980ja,Grunberg:1982fw}\footnote{FAC is inspired from Grunberg's idea of effective charge.}) stem from physical pictures: the result should be stable, the results should be maximally conformal, the convergence should be as fast as possible or, like in our case, the result should exhibit no over-subtraction of collinear singularities inside the PDFs.  The question is then how much one can depart from this expectation? Presumably, the scale value from a given prescription will not be the same at NLO and NNLO for instance. One could then try to derive the NLO scale from its formal expression artificially corrected by a typical NNLO corrections scaling like $\alpha_s/\pi$, on the order of unity in our case. This would provide a range of prescribed scales that we could plug in the NLO computation to get a scale uncertainty. In our case, instead of forcing $\log\frac{M_{\Q}^2}{\mu_F^2} +A_{\gamma i}$ to vanish, one could consider a range of $\mu_F$ such that its absolute value is bounded below unity. As such, we would probe the robustness of the scale-fixing criteria against expected higher-order corrections. This seems a very elegant solution for which the range of $\mu_F$ is $1/\sqrt{e} \leq \mu_F/\hat \mu_F \leq \sqrt{e}$. 

There is however a caveat to compare it with the usual way, not because of the above solution, but because the conventional method of scale variation by 2 is purely arbitrary\footnote{This however is not crucial when ones looks at how the scale uncertainties decrease with the order in $\alpha_S$. Yet, the absolute size of the uncertainty has essentially no physical meaning: a variation by $\sqrt{2}$, or 3, or whatever else not far from 2, would be equally acceptable while yielding different uncertainty estimates. We are in fact surprised how much this issue is underdiscussed when theory is confronted to experimental data.}. 
For meaningful comparisons, the range of {\it variation} should be accounted for. After all, what one looks after is the scale {\it dependence}. A natural way to proceed should instead be to compute $d\sigma/d\ln\mu$ which would then be a local estimation of the scale dependence. Assuming that the variation by 2 is an approximate way to numerically evaluate this derivative, the connection between both requires to include a $\ln 2$ factor and $\ln 2/\ln{e}$ for the proposal above. We refer to \ref{sec:scale} for more details.

In this context, we will show a NLO $\mu_F$ scale uncertainty derived from $ \ln{2} \times d\sigma/d\ln\mu$, in fact very similar to that obtained from $1/\sqrt{e} \leq \mu_F/\hat \mu_F \leq \sqrt{e}$ rescaled by $\ln 2/\ln{e}$.

%
% \vspace*{-0.2cm}
\section{\label{sec:results} Results}
Having discussed our methodology, let us now present and analyse our results for $J/\psi$ and $\Upsilon(1S)$ photoproduction cross sections computed at NLO with the $\hat{\mu}_F$ prescription.

\begin{figure}[hbt!]
    \centering
             \captionsetup[subfloat]{captionskip=-0.4cm,margin={0cm,0.8\columnwidth}}
\subfloat[]{\includegraphics[width=0.8\columnwidth]{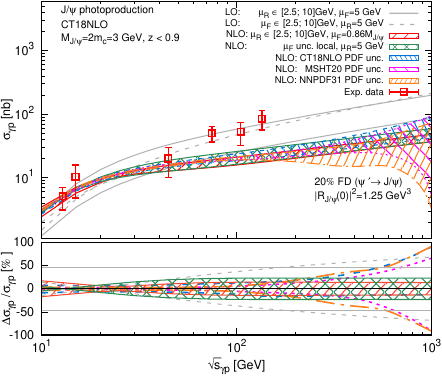}
    \label{fig:Jpsi_gammap}}\vspace*{-0.4cm} \\
\subfloat[]{\includegraphics[width=0.8\columnwidth]{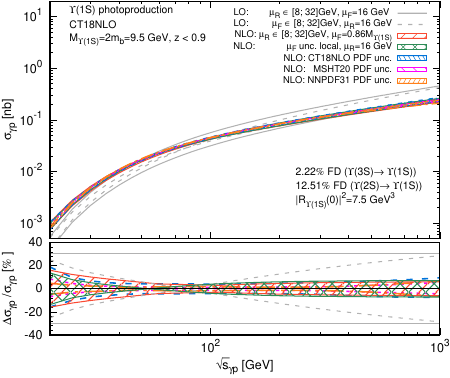}\label{fig:Y_gammap}}\vspace*{-0.5cm}
\caption{(Upper panels) $\sigma_{\gamma p}$ dependence on $\sqrt{s_{\gamma p}}$, (lower panels) $\Delta\sigma_{\gamma p}/\sigma_{\gamma p}$ dependence on $\sqrt{s_{\gamma p}}$ for (a) $J/\psi$ and (b) $\Upsilon(1S)$ inclusive photoproduction with the $\mu_R$ and the PDF uncertainties}\vspace*{-0.25cm} 
\end{figure} 

On \cf{fig:Jpsi_gammap} and \cf{fig:Y_gammap}, we have plotted (upper panel)  the cross sections $\sigma_{\gamma p}$ and (lower panel) its (scale and PDF)\footnote{We note here that the mass and $R(0)$ uncertainites are highly kinematically correlated and essentially translate into a quasi global offset. This thus why we focus on the $\mu_R$, $\mu_F$ and PDF uncertainties.} relative uncertainty  as functions of $\sqrt{s_{\gamma p}}$ for respectively $J/\psi$ and $\Upsilon(1S)$ photoproduction for different PDF sets: CT18NLO~\cite{Hou:2019efy}, MSHT20nlo\_as118~\cite{Bailey:2020ooq}, NNPDF31\_nlo\_as\_0118\_hessian~\cite{NNPDF:2017mvq}. Let us first discuss \cf{fig:Jpsi_gammap}. The LO cross section (2 solid and 2 dashed grey lines) relatively well describes the experimental data points in red with somewhat large uncertainties at large $\sqrt{s_{\gamma p}}$. The NLO cross section is systematically smaller and one notes that the NLO $\mu_R$ uncertainty (red hatched band) is reduced compared to the LO one, as expected from \cf{fig:mu_R/M}. The NLO $\mu_F$ uncertainty we obtained is well-behaved without any negative values anymore and is smaller than the LO one at large $\sqrt{s_{\gamma p}}$. 

The PDF uncertainties at NLO from CT18NLO, MSHT20nlo\_as118 and NNPDF31\_nlo\_as\_0118\_hessian are shown by respectively the blue, magenta and orange hatched bands. At large $\sqrt{s_{\gamma p}}$, which corresponds to the low-$x$ region in the proton, they naturally grow and eventually become larger than the $\mu_R$ uncertainty. Even though it is not an observable physical quantity, we note that with our present set-up (scheme and scale choice) the relative contribution from the $\gamma q$ fusion channel is relatively constant and close to 5\% from 20 GeV and above, about 95\% then comes from $\gamma g$ fusion. The three PDFs we have chosen are representative of what is available from fixed-order analyses. However, they are known~\cite{xFitterDevelopersTeam:2018hym} to be artificially suppressed at low $x$ and low scales and can even show a local mininum which then distorts in the energy dependence of $\eta_c$ hadroproduction cross section~\cite{Lansberg:2020ejc}. As such, the possibility remains, that even within their increasing uncertainties, such present PDF sets are possibly unsuited to reliably describe quarkonium production data. Using the latter to redetermine the former is then certainly something which should be attempted.

The increase of the PDF uncertainty is even more visible on the relative uncertainty plots \footnote{The LO relative scale uncertainties are computed as $\pm$($\sigma_{\gamma p}^{\rm max}-\sigma_{\gamma p}^{\rm min}$)/($2\sigma_{\gamma p}^{\rm cen}$), where $\sigma_{\gamma p}^{\rm max/min}$ is the maximum/minimum values of $\sigma_{\gamma p}$ obtained by varying $\mu_R$, $\mu_F$ in the quoted range and $\sigma_{\gamma p}^{\rm cen}$ is the cross section evaluated with $\mu_F = \mu_R = \mu_0$, with $\mu_0 = 5(16)$ GeV for $J/\psi$($\Upsilon(1S)$). At NLO, the $\mu_R$ uncertainty is calculated as $\pm$($\sigma_{\gamma p}^{\rm max}-\sigma_{\gamma p}^{\rm min}$)/($\sigma_{\gamma p}^{\rm max}+\sigma_{\gamma p}^{\rm min}$), while the $\mu_F$ uncertainty is estimated with the local method (see Sec.~\ref{sec:muf-local-unc} and \ref{sec:scale}), normalised to $\sigma_{\gamma p}^{\rm cen}(\mu_F = \hat{\mu}_F, \mu_R = \mu_0)$. For the PDF uncertainties, we used the normalised upper and lower PDF uncertainties~\cite{Buckley:2014ana} for $\mu_R=\mu_0$, where again $\mu_0 = 5 (16)$ GeV for $J/\psi$ ($\Upsilon(1S)$). }
\cf{fig:Jpsi_gammap} (lower panel) for $\mu_R=5$ GeV. Above 300 GeV, these are clearly larger than the $\mu_R$ one which slightly grows above 50~GeV due to the on-set of the negative contributions from the loop corrections (see below) and the $\mu_F$ one which also slightly grows above 20~GeV, where it happens to coincidentally vanish for a wide range of values.
As for the $\Upsilon(1S)$ case, shown on \cf{fig:Y_gammap}, the reduction of the $\mu_R$ uncertainty at NLO is further pronounced while the PDF and $\mu_F$ uncertainties remain similar.

\begin{figure}[h!]
    \centering
        \captionsetup[subfloat]{captionskip=-0.4cm,margin={0.8\columnwidth,0cm}}
\subfloat[]{\includegraphics[width=0.8\columnwidth]{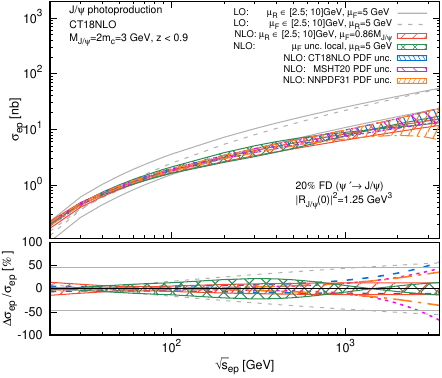}\label{fig:JPsi_ep}}\vspace*{-0.25cm}\\
\subfloat[]{\includegraphics[width=0.8\columnwidth]{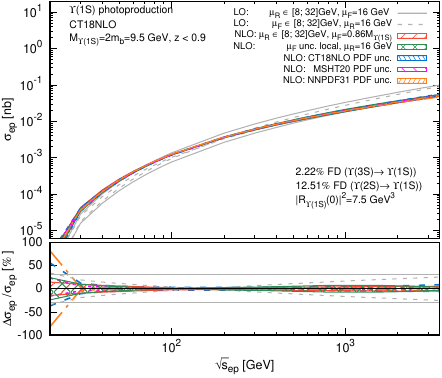}\label{fig:Y_ep}}\vspace*{-0.5cm}\\
\caption{(Upper panels) $\sigma_{e p}$  as a function of $\sqrt{s_{e p}}$, (lower panels) $\Delta\sigma_{e p}/\sigma_{e p}$ as a function of $\sqrt{s_{e p}}$ for (a) $J/\psi$ and (b) $\Upsilon(1S)$ inclusive photoproduction with its $\mu_R$ and the PDF uncertainties.\vspace*{-0.25cm}}
\end{figure}

In the $J/\psi$ case, it is clear that it will be important to have at our disposal computations at NNLO accuracy. As Kr\"amer noted~\cite{Kramer:1995nb} long ago, the ``virtual+soft'' contributions, encapsulated in {$c^{(1)}$},  are significantly more negative than for open heavy-flavour production~\cite{Smith:1991pw}. He suggested that this destructive interference with the Born order amplitude could be due to the momentum transfer of the exchanged virtual gluon, more likely to scatter the $Q\bar Q$ pair outside the static limit (${\bf p}\simeq 0$). {At NNLO, these one-loop amplitudes will be squared, the two-loop amplitudes will interfere with the Born amplitudes and the amplitudes of the one-loop corrections to the real-emission graphs will also interfere with the real-emission amplitudes. Unless the latter two are subject to the same strong destructive interference effect, one might expect relatively large positive NNLO corrections bringing the cross section  close to the upper limit of the LO range and then in better agreement with existing, yet old, data.}

At NNLO, we also expect a further reduction of the $\mu_R$ \new{and $\mu_F$}\footnote{It is legitimate to expect the oversubtraction by the AP counter terms to be reduced at NNLO associated with a reduction of the sensitivity on $\mu_F$. It might also be less sensitive on the PDF shape~\cite{Ozcelik:2019qze,Lansberg:2020ejc}.} uncertainties. This is particularly relevant especially around $50-100$ GeV, which corresponds to the EIC region. This would likely allow us to better probe gluon PDFs using photoproduction data. Going further, differential measurements in the elasticity or the rapidity could provide a complementary {leverage} in $x$ to fit the gluon PDF, even in the presence of sub-leading $v$ colour-octet contributions. Indeed, these would likely exhibit a very similar dependence on $x$. As we will see now, the expected yields at future facilities, in particular for charmonia, are clearly large enough to perform such differential measurements.

Let us now look at electron-proton cross sections as functions of $\sqrt{s_{ep}}$ for $J/\psi$ (\cf{fig:JPsi_ep}) and $\Upsilon(1S)$ photoproduction (\cf{fig:Y_ep}). To obtain them, \ce{eq:sigma_scaling_functions} was convoluted with the corresponding proton PDFs and a photon flux from the electron. We have used the same photon flux as in~\cite{Flore:2020jau}.

On \cf{fig:JPsi_ep} and \cf{fig:Y_ep}, the same colour code and the same parameters as for \cf{fig:Jpsi_gammap} and \cf{fig:Y_gammap} {have been used}. For $\sigma_{ep}$, one can see the same trends for the $\mu_R$ and PDF uncertainties as for $\sigma_{\gamma p}$. It is only at the LHeC energies and above that one could expect to constrain better the PDF uncertainty with such total cross section measurements unless we have at our disposal NNLO computations with yet smaller scale uncertainties.

\begin{table}[htbp!]\renewcommand{\arraystretch}{1}
    \begin{centering}\small
    \begin{tabular}{c|cccc}
    
        Exp. 
        & $\sqrt{s_{ep}}$
        & ${\cal L}$~(fb$^{-1})$
        & $N_{J/\psi}$ %
        & $N_{\Upsilon(1S)}$ %
        \\
   \hline
        EicC & 16.7 & 100 & $1.5^{+0.3}_{-0.2} \cdot 10^6$%
        &$2.3^{+1.1}_{-1.4} \cdot 10^0$
\\
        AMBER & 17.3 & {1} & $1.6^{+0.3}_{-0.3} \cdot 10^4$%
        &$<1$
\\
        EIC & 45 & 100 & $8.5^{+0.5}_{-1.0} \cdot 10^6$%
        & $6.1^{+0.7}_{-0.8} \cdot 10^2$
\\
        EIC &  140 & 100 & $2.5^{+0.1}_{-0.4} \cdot 10^7$%
        & $7.6^{+0.3}_{-0.7} \cdot 10^3$
\\
        LheC & 1183 & 100 & $9.3^{+2.9}_{-2.9} \cdot 10^7$%
        & $8.1^{+0.4}_{-0.7} \cdot 10^4$
\\
        FCC-eh & 3464 & 100 & $1.6^{+0.2}_{-1.0} \cdot 10^8$%
        & $1.8^{+0.1}_{-0.2} \cdot 10^5$
\\

    \end{tabular}
    \caption{Expected number of detected quarkonia at NLO  at different $\sqrt{s_{ep}}$ (in GeV) corresponding to future facilities
    (using CT18NLO, {$\mu_R=5$~GeV for $J/\psi$ and $\mu_R=16$~GeV for $\Upsilon(1S)$}, $\mu_F=\hat \mu_F$) for $\epsilon_{detect}=85\%$  {via} the decay channels to $\mu^+\mu^-$ and $e^+e^-$, namely $\epsilon^{J/\psi}_{ \ell^+ \ell^-}\approx0.1$, %
    and $\epsilon^{\Upsilon(1S)}_{ \ell^+ \ell^-}\approx0.04$. 
   }
    \label{tab:numb_of_particles} 
    \end{centering}\vspace*{-0.25cm}
\end{table}

In \ct{tab:numb_of_particles}, we provide estimations of the expected number of {$J/\psi$ and $\Upsilon(1S)$} possibly detected at the different $ep$  {c.m.} energies of planned experiments. As it can be seen, the expected %
{yields} are always very large for {$J/\psi$} which will clearly allow for a number of differential measurements in $z$, $y$ or $\sqrt{s_{\gamma p}}$. These could
{then} be used to reduce the impact of partially correlated theoretical uncertainties, from the scales and the heavy-quark mass affecting these photoproduction cross sections, in order to bring about some additional constraints on the PDFs at low scales, in particular the gluon one. {For $\Upsilon(1S)$, the yields should be sufficient to extract cross sections at the EIC, LHeC and FCC-eh even below their nominal luminosities. 

One can {also} estimate the expected number of detected $\psi',~\Upsilon(2S),~\Upsilon(3S)$ %
{using the following relations}  $N_{\psi'}\simeq0.07 \times  N_{J/\psi}$,$ N_{\Upsilon(2S)}\simeq0.4 \times N_{\Upsilon(1S)}$,
$N_{\Upsilon(3S)}\simeq0.3 \times N_{\Upsilon(1S)},$
{derived from the values of}\footnote{The relation for $\psi'$ was estimated using $|R_{\psi'}(0)|^2 = 0.8$ GeV$^3$.}} $|R_{\Q}(0)|^2$ and of the branching fractions to leptons. %
Using the above relations and the values in \ct{tab:numb_of_particles}, one can see that the yield of $\psi'$ should be measurable everywhere and the yields of $\Upsilon(2S)$ and $\Upsilon(3S)$ {are} close to %
about half of that of $\Upsilon(1S)$ {and} should be measurable at the EIC, LHeC and FCC-eh. The proximity between the $\Upsilon(nS)$ yields follows from their similar $|R_{\Q}(0)|^2$ and leptonic branchings.

\vspace*{-0.2cm}
\section{\label{sec:conclusions}A note on $P_T$-differential cross sections}

As announced, the present study focuses on the fully $P_T$-integrated case. If, instead, one is interested in $P_T$-differential cross sections, our prescription for a single scale process would not work as it stands. Indeed, one has to be cautious that a class of real-emission NLO corrections (\cf{init_gl_gl}) is kinematically enhanced by $(P_T/M_\Q)^2$ with respect to the Born contributions. Considering the expected $P_T$ behaviour of the scaling functions entering the definition of $\hat\mu_F$, one finds that $c_{\gamma i}^{(1)}$ (which contains these contributions) is enhanced with respect to $\overline{c}^{(1)}_{\gamma i}$ (which only arises from the collinear contributions). 

This would result in $\hat\mu_F(P_T)$ scaling as $M_\Q\times\exp(P_T/M_\Q)$. This can easily be explained if one reminds that our scale prescription effectively amounts to recast, at large partonic energies, all the NLO corrections in the PDF folded with the Born cross section. This would then be done, at large $P_T$, by trying to get those $P_T$-enhanced contributions entirely from the PDF evolution via an unphysically large $\mu_F$. At large $P_T$, the natural scale should instead be on the order of $P_T$ and not $M_\Q\times \exp(P_T/M_\Q)$.

Common dynamical scale choices for $P_T$-differential cross sections are $(0.5,1,2) \times m_T$ with $m_T=\sqrt{M_\Q^2+P_T^2}$. Two possible choices scaling like $m_T$  at large $P_T$ and compatible with $\hat\mu_F$ when integrated over $P_T$ would be (i) $\alpha \times  m_T$ with $\alpha$ fixed such that $\hat\mu_F=\alpha \sqrt{M_\Q^2+\langle P_T^2 \rangle}$
and (ii) $ \sqrt{(\beta  \times M_\Q)^2+P_T^2 }$ with $\beta$ fixed such that  $\hat\mu_F= \sqrt{(\beta M_\Q)^2+\langle P_T^2 \rangle}$. At HERA energies~\cite{Aid:1996dn}, $\langle P_T^2 \rangle \simeq 2.5$~GeV$^2$ for $J/\psi$, this gives $\alpha=0.77$ and $\beta =0.7$. The former choice is shown on \cf{fig:dsigdpt} (red boxes) and compared to cross section obtained with a fixed $\mu_F=\hat\mu_F$ for different $\mu_R$ (hashed red histogram). All choices give similar results which are compatible with the latest H1 data~\cite{Aaron:2010gz}. Predictions for the EIC at 140 GeV are also given. They confirm predictions given in our previous study~\cite{Flore:2020jau} with an approximate NLO computation.

\begin{figure}[t]
\centering\includegraphics[width=0.85\columnwidth]{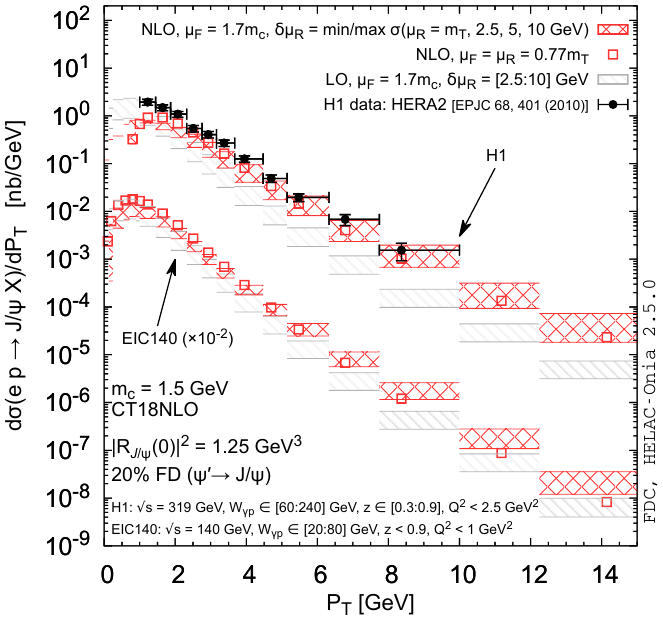}\vspace*{-0.4cm} 
\caption{LO and NLO $d\sigma_{ep}/dP_T$ as a function of $P_T$ for $J/\psi$ photoproduction for different scale choices at HERA compared with the latest H1 experimental data~\cite{Aaron:2010gz} and at the EIC140.
}
\label{fig:dsigdpt}
\end{figure}

%
% \vspace*{-0.2cm}
\section{\label{sec:conclusions}Conclusions}

%In this work, we have revisited the inclusive photoproduction up to NLO for $J/\psi$ and $\Upsilon(1S)$ at lepton-proton colliders. To this end, we have computed the $P_T$- and $z$-integrated $\sigma_{ep}$ and $\sigma_{\gamma p}$.

Like for other charmonium production processes~\cite{Mangano:1996kg,Feng:2015cba,Lansberg:2020ejc}, we have observed the appearance at NLO of negative total cross section which we attribute to an oversubtraction of collinear divergences into the PDF via AP-CT in the $\msbar$ scheme. We applied the $\hat \mu_F$ {prescription} proposed in \cite{Lansberg:2020ejc}, {which up to NLO corresponds to a resummation of such collinear divergences in  High-Energy Factorisation (HEF)~\cite{Lansberg:2021vie}}. 
Expressing this integrated cross section in terms of scaling functions exhibiting its explicit $\mu_R$ and $\mu_F$ scale dependencies, we have found that, for $z<0.9$,  {the optimal factorisation scale is} $\hat \mu_F={0.86} M_\Q$ which falls well within the usual ranges of used values. {Like} for $\eta_c$ hadroproduction, such a factorisation-scale prescription indeed allows one to avoid negative NLO cross sections, but it apparently prevents one from studying the corresponding factorisation-scale uncertainties. We have thus elaborated on two approaches to study such scale uncertainties when one scale is fixed by a physical argument. We were not aware of such ideas before and these clearly deserve dedicated studies in the future.

We have seen that the NLO $\mu_R$ uncertainties {get} reduced compared to the LO ones but slightly increase around 50~GeV, because of {rather large} (negative\footnote{{Let us stress that unless $\mu_R$ is taken very small with a large $\alpha_s(\mu_R)$, these negative contributions are not problematic, unlike the oversubtraction by the AP-CT.}}) interferences between the one-loop and Born amplitudes.
As mentioned before, these "virtual+soft" contributions are significantly more negative than for open heavy-flavour production.  While Kr\"amer suggested that the difference could stem from the static limit ($\mathbf{p}\simeq 0$) specific to the non-relativistic quarkonia, from which one easily departs when gluon exchanges occur, it will certainly be very instructive to have NNLO computations to see whether such one-loop amplitudes squared would bring the cross section back up close to the LO one or whether the interference between the two-loop and the Born amplitudes and between the real-virtual and the real amplitudes would be also negative and large. 

Our evaluated $\mu_F$ uncertainties at NLO are also reduced compared to those at LO at large $\sqrt{s_{\gamma p}}$ which were totally out of control before the application of the $\hat \mu_F$ {prescription}. We see that as a very encouraging sign.

In any case, at NNLO, it is reasonable to expect a further reduction of the scale uncertainties compared to the NLO results.
{We have also briefly addressed the application of our scale prescription to $P_T$-differentical cross sections and proposed choices scaling like  $m_T$ at large $P_T$ while compatible with $\hat \mu_F={0.86} M_\Q$ when $P_T$ is integrated over.}

We have also qualitatively investigated the possibility to constrain PDFs using future $J/\psi$ and $\Upsilon(1S)$ photoproduction data. Indeed, present PDF sets are possibly~\cite{Lansberg:2020ejc} unsuited to reliably describe high-energy quarkonium production data. Restricting to conventional sets, we have seen that PDF uncertainties get larger than the (NLO) $\mu_R$ and $\mu_F$ uncertainties with the growth of the $\gamma p$ c.m.~energy, in practice from around $300$ GeV, \ie ~for $x$ below $0.01$.  Although this is above the reach of the future EIC, we hope that with NNLO predictions at our disposal in the future, with yet smaller $\mu_R$ and $\mu_F$ uncertainties, one could set novel constraints on PDFs with such EIC measurements.  Given our estimated counting rates for 100~fb$^{-1}$ of $ep$ collisions, we expect that a number of differential measurements will be possible to reduce the impact of highly or even partially correlated theoretical uncertainties, including the contamination of higher-$v$ corrections such as the colour-octet contributions. These, along with the forthcoming HL-LHC measurements~\cite{Chapon:2020heu}, should also definitely  help to improve our understanding of the quarkonium-production mechanisms.

Strictly speaking our predictions for $J/\psi$ and $\psi'$ only regard the prompt yields. An evaluation at NLO of the beauty production cross section points at a feed-down fraction at the 5\% level. Given the larger size of the other uncertainties and the possibility to remove it experimentally, we have neglected it. In general though, it will be useful to have a dedicated experimental measurements at the EIC at least to measure the beauty feed down. It may become more significant at low $z$ where the resolved-photon contribution could set in at high $\sqrt{s_{ep}}$, like we have seen~\cite{Flore:2020jau} it to become the dominant source of $J/\psi$ at large $P_T$.

% \vspace*{-0.85cm}
\paragraph*{\bf Acknowledgements}

We thank S.J.~Brodsky, Y.~Dokshitzer, M.~Mangano, M.~Nefedov and H.~Sazdjian for useful discussions, L.~Manna for a NLO estimate of the $b$ photoproduction cross section and K.~Lynch for comments on the manuscript.
This project has received funding from the European Union's Horizon 2020 research and innovation programme under grant agreement No.~824093 in order
to contribute to the EU Virtual Access {\sc NLOAccess}.
This project has also received funding from the Agence Nationale de la Recherche (ANR) via the grant ANR-20-CE31-0015 (``PrecisOnium'') and via the IDEX Paris-Saclay "Investissements d’Avenir" (ANR-11-IDEX-0003-01) through the GLUODYNAMICS project funded by the "P2IO LabEx (ANR-10-LABX-0038)" and through the Joint PhD Programme of Universit\'e Paris-Saclay (ADI).
This work  was also partly supported by the French CNRS via the IN2P3 project GLUE@NLO, via the Franco-Chinese LIA FCPPL (Quarkonium4AFTER) and via the Franco-Polish EIA (GlueGraph).
M.A.O.’s work was partly supported by the ERC grant 637019 “MathAm”.
C.F.~has been also partially supported by Fondazione di Sardegna under the project “Proton tomography at the LHC”, project number F72F20000220007 (University of Cagliari), and is thankful to the Physics Department of Cagliari University for the hospitality and support for his visit, during which part of the project was done.

\vspace*{-0.25cm}
\appendix\vspace*{-0.25cm}
\renewcommand{\thefigure}{\arabic{figure}}
\section{\label{sec:LDME} Leptonic width and wave function at the origin}

Up to NNLO, one has~\cite{Czarnecki:1997vz,Beneke:1997jm} (for $\mu_{\rm NRQCD}=m_Q$):
\begin{equation}
\begin{aligned}
\Gamma_{\ell\ell}&=\frac{4\pi\alpha^2 e_Q^2 f_{\Q}^2}{3M_{\Q}},f_{\Q}=\sqrt{\frac{3}{\pi M_{\Q}}|R(0)|^2} \Bigg[1 - \frac{8}{3}\frac{\alpha_s(\mu_R)}{\pi}  \\
&-\Biggl(44.55-0.41n_{lf} -\frac{2}{3}\beta_0\ln{\frac{M_{\Q}^2}{4\mu_R^2}}\Biggr)\Biggl(\frac{\alpha_s(\mu_R)}{\pi}\Biggr)^2
\Bigg],
\label{eq_decaycorr}
\end{aligned}
\end{equation}
\noindent where $n_{lf}$ is the number of active light flavours, $\alpha$ is the electromagnetic coupling constant, $\alpha_s$ is the strong interaction coupling, $M_{\Q}$ is the $\Q$ mass, $e_Q$ is the magnitude of the heavy-quark charge (in units of the electron charge).  In Table \ref{tab:ldme}, we have gathered the resulting radial part of the Schr\"odinger wave function at the origin of the configuration space at LO, NLO and NNLO for $J/\psi$ %
and $\Upsilon(1S)$, which were computed from \ce{eq_decaycorr} {with} the measured value of $\Gamma_{\ell\ell}$~\cite{ParticleDataGroup:2020ssz}. 
% \vspace*{-0.5cm}

\begin{table}[htbp]
    \begin{centering}\small\renewcommand{\arraystretch}{1.25}
    \begin{tabular}{c|c|c|c}
        $\alpha_{s}$ 
        & $|R_{J/\psi}(0)|^2_{\rm LO}$ 
        & $|R_{J/\psi}(0)|^2_{\rm NLO}$
        & $|R_{J/\psi}(0)|^2_{\rm NNLO}$
        \\
    \hline
           [0.18,0.34]
           &0.52
           &[0.73,1.04]
           &[1.32,12.61]
\\
     \hline \hline
        $\alpha_{s}$ 
        & $|R_{\Upsilon (1S)}(0)|^2_{\rm LO}$ 
        & $|R_{\Upsilon (1S)}(0)|^2_{\rm NLO}$
        & $|R_{\Upsilon (1S)}(0)|^2_{\rm NNLO}$
        \\
    \hline
           [0.14,0.18]
           &4.90
           &[6.31,6.83]
           &[8.63,10.38]
\\
    
    \end{tabular}
    \caption{Values of $|R(0)|^2$ in GeV$^3$ extracted from the corresponding leptonic widths~\cite{ParticleDataGroup:2020ssz}
    at LO, NLO and NNLO.
    }
    \label{tab:ldme}
    \end{centering}
\end{table}

\section{\label{sec:scale} Estimating the `scale' uncertainty for a fixed scale}

Scale uncertainties are historically evaluated in pQCD by a multiplicative variation by a factor of 2. Assuming that the corresponding $\mu$ uncertainty, defined as $\sigma \pm \Delta \sigma$, is evaluated via

\begin{equation}
    \Delta\sigma(\mu)=\frac{\left|\sigma(2\mu)-\sigma(\mu/2)\right|}{2}\,,
\end{equation}

\noindent it is easy to note that wider variations necessarily lead to larger uncertainties and narrower ones to smaller uncertainties. However, as pointed out above, the common practice is to use such a factor 2 which then lead to the usual 7- or 9-point variation technique to estimate the envelope corresponding to the factorisation and renormalisation scale uncertainty.

To `fix' this ambiguity and yet allowing one to make the variation as wide or narrow as we wish, we propose to rescale it as follows

\begin{equation}
\label{eq:delta_xi_sigma}
    \begin{aligned}
\Delta_\xi\sigma(\mu)=\left|\frac{\sigma(\xi\mu)-\sigma(\mu/\xi)}{2} \frac{\ln 2}{\ln\xi}\right|.
    \end{aligned}
\end{equation}
For $\xi=2$, one recovers the usual uncertainty.

With this definition, one can consider $\Delta_\xi\sigma$ evaluated \emph{locally} (\textit{i.e.}~for a scale variation very close to $\mu$, thus for $\xi \to 1$). Such local evaluation is in fact simply connected to $\frac{\partial \sigma}{\partial \ln \mu}$ as 
\footnote{We assume that the symmetric derivative equates the usual derivative.}
\begin{equation}
\lim_{\xi\to 1}\Delta_\xi\sigma  = \ln{2} \times \left|\frac{\partial \sigma (\mu)}{\partial \ln\mu}\right|\,.
\end{equation}

\bibliographystyle{utphys}
\biboptions{sort&compress}
% \bibliography{NLO-photoproduction_021022}

\providecommand{\href}[2]{#2}\begingroup\raggedright\begin{thebibliography}{10}

\bibitem{Lansberg:2019adr}
J.-P. Lansberg, ``{New Observables in Inclusive Production of Quarkonia},''
  \href{http://dx.doi.org/10.1016/j.physrep.2020.08.007}{{\em Phys. Rept.}
  {\bfseries 889} (2020) 1--106},
  \href{http://arxiv.org/abs/1903.09185}{{\ttfamily arXiv:1903.09185
  [hep-ph]}}.

\bibitem{Andronic:2015wma}
A.~Andronic {\em et~al.}, ``{Heavy-flavour and quarkonium production in the LHC
  era: from proton–proton to heavy-ion collisions},''
  \href{http://dx.doi.org/10.1140/epjc/s10052-015-3819-5}{{\em Eur. Phys. J.}
  {\bfseries C76} no.~3, (2016) 107},
\href{http://arxiv.org/abs/1506.03981}{{\ttfamily arXiv:1506.03981 [nucl-ex]}}.
%%CITATION = ARXIV:1506.03981;%%.

\bibitem{Brambilla:2010cs}
N.~Brambilla {\em et~al.}, ``{Heavy Quarkonium: Progress, Puzzles, and
  Opportunities},''
  \href{http://dx.doi.org/10.1140/epjc/s10052-010-1534-9}{{\em Eur. Phys. J.}
  {\bfseries C71} (2011) 1534},
\href{http://arxiv.org/abs/1010.5827}{{\ttfamily arXiv:1010.5827 [hep-ph]}}.
%%CITATION = ARXIV:1010.5827;%%.

\bibitem{Lansberg:2006dh}
J.~P. Lansberg, ``{$J/\psi$, $\psi'$ and $\Upsilon$ production at hadron
  colliders: A Review},''
  \href{http://dx.doi.org/10.1142/S0217751X06033180}{{\em Int. J. Mod. Phys.}
  {\bfseries A21} (2006) 3857--3916},
\href{http://arxiv.org/abs/hep-ph/0602091}{{\ttfamily arXiv:hep-ph/0602091
  [hep-ph]}}.
%%CITATION = HEP-PH/0602091;%%.

\bibitem{Brambilla:2004wf}
{\bfseries Quarkonium Working Group} Collaboration, N.~Brambilla {\em et~al.},
  ``{Heavy quarkonium physics},''
\href{http://arxiv.org/abs/hep-ph/0412158}{{\ttfamily arXiv:hep-ph/0412158
  [hep-ph]}}.
%%CITATION = HEP-PH/0412158;%%.

\bibitem{Kramer:2001hh}
M.~Kraemer, ``{Quarkonium production at high-energy colliders},''
  \href{http://dx.doi.org/10.1016/S0146-6410(01)00154-5}{{\em Prog. Part. Nucl.
  Phys.} {\bfseries 47} (2001) 141--201},
\href{http://arxiv.org/abs/hep-ph/0106120}{{\ttfamily arXiv:hep-ph/0106120
  [hep-ph]}}.
%%CITATION = HEP-PH/0106120;%%.

\bibitem{Flore:2020jau}
C.~Flore, J.-P. Lansberg, H.-S. Shao, and Y.~Yedelkina, ``{Large-$P_T$
  inclusive photoproduction of $J/\psi$ in electron-proton collisions at HERA
  and the EIC},'' \href{http://dx.doi.org/10.1016/j.physletb.2020.135926}{{\em
  Phys. Lett. B} {\bfseries 811} (2020) 135926},
  \href{http://arxiv.org/abs/2009.08264}{{\ttfamily arXiv:2009.08264
  [hep-ph]}}.

\bibitem{Flore:2021rlc}
C.~Flore, J.-P. Lansberg, H.-S. Shao, and Y.~Yedelkina, ``{NLO inclusive $J/\psi$
  photoproduction at large $P_T$ at HERA and the EIC},''
  \href{http://dx.doi.org/10.21468/SciPostPhysProc.8.011}{{\em SciPost Phys.
  Proc.} {\bfseries 8} (2022) 011},
  \href{http://arxiv.org/abs/2107.13434}{{\ttfamily arXiv:2107.13434
  [hep-ph]}}.

\bibitem{Chang:1979nn}
C.-H. Chang, ``{Hadronic Production of $J/\psi$ Associated With a Gluon},''
\href{http://dx.doi.org/10.1016/0550-3213(80)90175-3}{{\em Nucl. Phys.}
  {\bfseries B172} (1980) 425--434}.
%%CITATION = NUPHA,B172,425;%%.

\bibitem{Berger:1980ni}
E.~L. Berger and D.~L. Jones, ``{Inelastic Photoproduction of J/psi and Upsilon
  by Gluons},''
\href{http://dx.doi.org/10.1103/PhysRevD.23.1521}{{\em Phys. Rev.} {\bfseries
  D23} (1981) 1521--1530}.
%%CITATION = PHRVA,D23,1521;%%.

\bibitem{Baier:1983va}
R.~Baier and R.~Ruckl, ``{Hadronic Collisions: A Quarkonium Factory},''
\href{http://dx.doi.org/10.1007/BF01572254}{{\em Z. Phys.} {\bfseries C19}
  (1983) 251}.
%%CITATION = ZEPYA,C19,251;%%.

\bibitem{Bodwin:1994jh}
G.~T. Bodwin, E.~Braaten, and G.~P. Lepage, ``{Rigorous QCD analysis of
  inclusive annihilation and production of heavy quarkonium},''
  \href{http://dx.doi.org/10.1103/PhysRevD.55.5853,
  10.1103/PhysRevD.51.1125}{{\em Phys. Rev.} {\bfseries D51} (1995)
  1125--1171}, \href{http://arxiv.org/abs/hep-ph/9407339}{{\ttfamily
  arXiv:hep-ph/9407339 [hep-ph]}}.
[Erratum: Phys. Rev.D55,5853(1997)].
%%CITATION = HEP-PH/9407339;%%.

\bibitem{Aid:1996dn}
{\bfseries H1} Collaboration, S.~Aid {\em et~al.}, ``{Elastic and inelastic
  photoproduction of $J/\psi$ mesons at HERA},''
  \href{http://dx.doi.org/10.1016/0550-3213(96)00274-X}{{\em Nucl. Phys. B}
  {\bfseries 472} (1996) 3--31},
  \href{http://arxiv.org/abs/hep-ex/9603005}{{\ttfamily arXiv:hep-ex/9603005}}.

\bibitem{Breitweg:1997we}
{\bfseries ZEUS} Collaboration, J.~Breitweg {\em et~al.}, ``{Measurement of
  inelastic $J/\psi$ photoproduction at HERA},''
  \href{http://dx.doi.org/10.1007/s002880050583}{{\em Z. Phys. C} {\bfseries
  76} (1997) 599--612}, \href{http://arxiv.org/abs/hep-ex/9708010}{{\ttfamily
  arXiv:hep-ex/9708010}}.

\bibitem{Chekanov:2002at}
{\bfseries ZEUS} Collaboration, S.~Chekanov {\em et~al.}, ``{Measurements of
  inelastic $J /\psi$ and $\psi^\prime$ photoproduction at HERA},''
  \href{http://dx.doi.org/10.1140/epjc/s2002-01130-2}{{\em Eur. Phys. J. C}
  {\bfseries 27} (2003) 173--188},
  \href{http://arxiv.org/abs/hep-ex/0211011}{{\ttfamily arXiv:hep-ex/0211011}}.

\bibitem{Adloff:2002ex}
{\bfseries H1} Collaboration, C.~Adloff {\em et~al.}, ``{Inelastic
  photoproduction of $J/\psi$ mesons at HERA},''
  \href{http://dx.doi.org/10.1007/s10052-002-1009-8}{{\em Eur. Phys. J.}
  {\bfseries C25} (2002) 25--39},
\href{http://arxiv.org/abs/hep-ex/0205064}{{\ttfamily arXiv:hep-ex/0205064
  [hep-ex]}}.
%%CITATION = HEP-EX/0205064;%%.

\bibitem{Chekanov:2009ad}
{\bfseries ZEUS} Collaboration, S.~Chekanov {\em et~al.}, ``{Measurement of
  J/psi helicity distributions in inelastic photoproduction at HERA},''
  \href{http://dx.doi.org/10.1088/1126-6708/2009/12/007}{{\em JHEP} {\bfseries
  12} (2009) 007},
\href{http://arxiv.org/abs/0906.1424}{{\ttfamily arXiv:0906.1424 [hep-ex]}}.
%%CITATION = ARXIV:0906.1424;%%.

\bibitem{Aaron:2010gz}
{\bfseries H1} Collaboration, F.~Aaron {\em et~al.}, ``{Inelastic Production of
  $J/\psi$ Mesons in Photoproduction and Deep Inelastic Scattering at HERA},''
  \href{http://dx.doi.org/10.1140/epjc/s10052-010-1376-5}{{\em Eur. Phys. J. C}
  {\bfseries 68} (2010) 401--420},
  \href{http://arxiv.org/abs/1002.0234}{{\ttfamily arXiv:1002.0234 [hep-ex]}}.

\bibitem{Abramowicz:2012dh}
{\bfseries ZEUS} Collaboration, H.~Abramowicz {\em et~al.}, ``{Measurement of
  inelastic $J/\psi$ and $\psi^\prime$ photoproduction at HERA},''
  \href{http://dx.doi.org/10.1007/JHEP02(2013)071}{{\em JHEP} {\bfseries 02}
  (2013) 071},
\href{http://arxiv.org/abs/1211.6946}{{\ttfamily arXiv:1211.6946 [hep-ex]}}.
%%CITATION = ARXIV:1211.6946;%%.

\bibitem{Jung:1992uj}
H.~Jung, G.~A. Schuler, and J.~Terron, ``{J / psi production mechanisms and
  determination of the gluon density at HERA},''
\href{http://dx.doi.org/10.1142/S0217751X92003604}{{\em Int. J. Mod. Phys.}
  {\bfseries A7} (1992) 7955--7988}.
%%CITATION = IMPAE,A7,7955;%%.

\bibitem{Chapon:2020heu}
E.~Chapon {\em et~al.}, ``{Prospects for quarkonium studies at the
  high-luminosity LHC},''
  \href{http://dx.doi.org/10.1016/j.ppnp.2021.103906}{{\em Prog. Part. Nucl.
  Phys.} {\bfseries 122} (2022) 103906},
  \href{http://arxiv.org/abs/2012.14161}{{\ttfamily arXiv:2012.14161
  [hep-ph]}}.

\bibitem{Boer:2020bbd}
D.~Boer, U.~D'Alesio, F.~Murgia, C.~Pisano, and P.~Taels, ``{$J/\psi$ meson
  production in SIDIS: matching high and low transverse momentum},''
  \href{http://dx.doi.org/10.1007/JHEP09(2020)040}{{\em JHEP} {\bfseries 09}
  (2020) 040},
\href{http://arxiv.org/abs/2004.06740}{{\ttfamily arXiv:2004.06740 [hep-ph]}}.
%%CITATION = ARXIV:2004.06740;%%.

\bibitem{Kishore:2019fzb}
R.~Kishore, A.~Mukherjee, and S.~Rajesh, ``{Sivers asymmetry in the
  photoproduction of a $J/\psi$ and a jet at the EIC},''
  \href{http://dx.doi.org/10.1103/PhysRevD.101.054003}{{\em Phys. Rev.}
  {\bfseries D101} no.~5, (2020) 054003},
\href{http://arxiv.org/abs/1908.03698}{{\ttfamily arXiv:1908.03698 [hep-ph]}}.
%%CITATION = ARXIV:1908.03698;%%.

\bibitem{DAlesio:2019qpk}
U.~D'Alesio, F.~Murgia, C.~Pisano, and P.~Taels, ``{Azimuthal asymmetries in
  semi-inclusive $J/\psi\,+\,\mathrm{jet}$ production at an EIC},''
  \href{http://dx.doi.org/10.1103/PhysRevD.100.094016}{{\em Phys. Rev.}
  {\bfseries D100} no.~9, (2019) 094016},
\href{http://arxiv.org/abs/1908.00446}{{\ttfamily arXiv:1908.00446 [hep-ph]}}.
%%CITATION = ARXIV:1908.00446;%%.

\bibitem{Bacchetta:2018ivt}
A.~Bacchetta, D.~Boer, C.~Pisano, and P.~Taels, ``{Gluon TMDs and NRQCD matrix
  elements in $J/\psi$ production at an EIC},''
  \href{http://dx.doi.org/10.1140/epjc/s10052-020-7620-8}{{\em Eur. Phys. J.}
  {\bfseries C80} no.~1, (2020) 72},
\href{http://arxiv.org/abs/1809.02056}{{\ttfamily arXiv:1809.02056 [hep-ph]}}.
%%CITATION = ARXIV:1809.02056;%%.

\bibitem{Kramer:1994zi}
M.~Kramer, J.~Zunft, J.~Steegborn, and P.~M. Zerwas, ``{Inelastic J / psi
  photoproduction},''
  \href{http://dx.doi.org/10.1016/0370-2693(95)00155-E}{{\em Phys. Lett. B}
  {\bfseries 348} (1995) 657--664},
  \href{http://arxiv.org/abs/hep-ph/9411372}{{\ttfamily arXiv:hep-ph/9411372}}.

\bibitem{Kramer:1995nb}
M.~Kraemer, ``{QCD corrections to inelastic J / psi photoproduction},''
  \href{http://dx.doi.org/10.1016/0550-3213(95)00568-4}{{\em Nucl. Phys.}
  {\bfseries B459} (1996) 3--50},
\href{http://arxiv.org/abs/hep-ph/9508409}{{\ttfamily arXiv:hep-ph/9508409
  [hep-ph]}}.
%%CITATION = HEP-PH/9508409;%%.

\bibitem{H1:1996kyo}
{\bfseries H1} Collaboration, S.~Aid {\em et~al.}, ``{Elastic and inelastic
  photoproduction of $J/\psi$ mesons at HERA},''
  \href{http://dx.doi.org/10.1016/0550-3213(96)00274-X}{{\em Nucl. Phys. B}
  {\bfseries 472} (1996) 3--31},
  \href{http://arxiv.org/abs/hep-ex/9603005}{{\ttfamily arXiv:hep-ex/9603005}}.

\bibitem{Lansberg:2020ejc}
J.-P. Lansberg and M.~A. Ozcelik, ``{Curing the unphysical behaviour of NLO
  quarkonium production at the LHC and its relevance to constrain the gluon PDF
  at low scales},''
  \href{http://dx.doi.org/10.1140/epjc/s10052-021-09258-7}{{\em Eur. Phys. J.
  C} {\bfseries 81} no.~6, (2021) 497},
  \href{http://arxiv.org/abs/2012.00702}{{\ttfamily arXiv:2012.00702
  [hep-ph]}}.

\bibitem{Accardi:2012qut}
A.~Accardi {\em et~al.}, ``{Electron Ion Collider: The Next QCD Frontier}:
  {Understanding the glue that binds us all},''
  \href{http://dx.doi.org/10.1140/epja/i2016-16268-9}{{\em Eur. Phys. J. A}
  {\bfseries 52} no.~9, (2016) 268},
  \href{http://arxiv.org/abs/1212.1701}{{\ttfamily arXiv:1212.1701 [nucl-ex]}}.

\bibitem{LHeCStudyGroup:2012zhm}
{\bfseries LHeC Study Group} Collaboration, J.~L. Abelleira~Fernandez {\em
  et~al.}, ``{A Large Hadron Electron Collider at CERN: Report on the Physics
  and Design Concepts for Machine and Detector},''
  \href{http://dx.doi.org/10.1088/0954-3899/39/7/075001}{{\em J. Phys. G}
  {\bfseries 39} (2012) 075001},
  \href{http://arxiv.org/abs/1206.2913}{{\ttfamily arXiv:1206.2913
  [physics.acc-ph]}}.

\bibitem{FCC:2018byv}
{\bfseries FCC} Collaboration, A.~Abada {\em et~al.}, ``{FCC Physics
  Opportunities}: {Future Circular Collider Conceptual Design Report Volume
  1},'' \href{http://dx.doi.org/10.1140/epjc/s10052-019-6904-3}{{\em Eur. Phys.
  J. C} {\bfseries 79} no.~6, (2019) 474}.

\bibitem{Adams:2018pwt}
B.~Adams {\em et~al.}, ``{Letter of Intent: A New QCD facility at the M2 beam
  line of the CERN SPS (COMPASS++/AMBER)},''
  \href{http://arxiv.org/abs/1808.00848}{{\ttfamily arXiv:1808.00848
  [hep-ex]}}.

\bibitem{Anderle:2021wcy}
D.~P. Anderle {\em et~al.}, ``{Electron-ion collider in China},''
  \href{http://dx.doi.org/10.1007/s11467-021-1062-0}{{\em Front. Phys.
  (Beijing)} {\bfseries 16} no.~6, (2021) 64701},
  \href{http://arxiv.org/abs/2102.09222}{{\ttfamily arXiv:2102.09222
  [nucl-ex]}}.

\bibitem{Artoisenet:2008fc}
P.~Artoisenet, J.~M. Campbell, J.-P. Lansberg, F.~Maltoni, and F.~Tramontano,
  ``{$\Upsilon$ Production at Fermilab Tevatron and LHC Energies},''
  \href{http://dx.doi.org/10.1103/PhysRevLett.101.152001}{{\em Phys. Rev.
  Lett.} {\bfseries 101} (2008) 152001},
  \href{http://arxiv.org/abs/0806.3282}{{\ttfamily arXiv:0806.3282 [hep-ph]}}.

\bibitem{Lansberg:2008gk}
J.-P. Lansberg, ``{On the mechanisms of heavy-quarkonium hadroproduction},''
  \href{http://dx.doi.org/10.1140/epjc/s10052-008-0826-9}{{\em Eur. Phys. J. C}
  {\bfseries 61} (2009) 693--703},
  \href{http://arxiv.org/abs/0811.4005}{{\ttfamily arXiv:0811.4005 [hep-ph]}}.

\bibitem{Lansberg:2009db}
J.-P. Lansberg, ``{Real next-to-next-to-leading-order QCD corrections to
  $J/\psi$ and Upsilon hadroproduction in association with a photon},''
  \href{http://dx.doi.org/10.1016/j.physletb.2009.07.067}{{\em Phys. Lett. B}
  {\bfseries 679} (2009) 340--346},
  \href{http://arxiv.org/abs/0901.4777}{{\ttfamily arXiv:0901.4777 [hep-ph]}}.

\bibitem{Gong:2012ah}
B.~Gong, J.-P. Lansberg, C.~Lorce, and J.~Wang, ``{Next-to-leading-order QCD
  corrections to the yields and polarisations of J/Psi and Upsilon directly
  produced in association with a Z boson at the LHC},''
  \href{http://dx.doi.org/10.1007/JHEP03(2013)115}{{\em JHEP} {\bfseries 03}
  (2013) 115},
\href{http://arxiv.org/abs/1210.2430}{{\ttfamily arXiv:1210.2430 [hep-ph]}}.
%%CITATION = ARXIV:1210.2430;%%.

\bibitem{Lansberg:2013qka}
J.-P. Lansberg and H.-S. Shao, ``{Production of $J/\psi + \eta_{c}$ versus
  $J/\psi + J/\psi$ at the LHC: Importance of Real $\alpha^{5}_{s}$
  Corrections},'' \href{http://dx.doi.org/10.1103/PhysRevLett.111.122001}{{\em
  Phys. Rev. Lett.} {\bfseries 111} (2013) 122001},
\href{http://arxiv.org/abs/1308.0474}{{\ttfamily arXiv:1308.0474 [hep-ph]}}.
%%CITATION = ARXIV:1308.0474;%%.

\bibitem{Lansberg:2014swa}
J.-P. Lansberg and H.-S. Shao, ``{J/psi -pair production at large momenta:
  Indications for double parton scatterings and large $\alpha_s^5$
  contributions},''
  \href{http://dx.doi.org/10.1016/j.physletb.2015.10.083}{{\em Phys. Lett.}
  {\bfseries B751} (2015) 479--486},
\href{http://arxiv.org/abs/1410.8822}{{\ttfamily arXiv:1410.8822 [hep-ph]}}.
%%CITATION = ARXIV:1410.8822;%%.

\bibitem{Lansberg:2017ozx}
J.-P. Lansberg, H.-S. Shao, and H.-F. Zhang, ``{$\eta_c'$ Hadroproduction at
  Next-to-Leading Order and its Relevance to $\psi'$ Production},''
  \href{http://dx.doi.org/10.1016/j.physletb.2018.10.009}{{\em Phys. Lett. B}
  {\bfseries 786} (2018) 342--346},
  \href{http://arxiv.org/abs/1711.00265}{{\ttfamily arXiv:1711.00265
  [hep-ph]}}.

\bibitem{Shao:2018adj}
H.-S. Shao, ``{Boosting perturbative QCD stability in quarkonium production},''
  \href{http://dx.doi.org/10.1007/JHEP01(2019)112}{{\em JHEP} {\bfseries 01}
  (2019) 112}, \href{http://arxiv.org/abs/1809.02369}{{\ttfamily
  arXiv:1809.02369 [hep-ph]}}.

\bibitem{Fritzsch:1977ay}
H.~Fritzsch, ``{Producing Heavy Quark Flavors in Hadronic Collisions: A Test of
  Quantum Chromodynamics},''
  \href{http://dx.doi.org/10.1016/0370-2693(77)90108-3}{{\em Phys. Lett. B}
  {\bfseries 67} (1977) 217--221}.

\bibitem{Halzen:1977rs}
F.~Halzen, ``{Cvc for Gluons and Hadroproduction of Quark Flavors},''
  \href{http://dx.doi.org/10.1016/0370-2693(77)90144-7}{{\em Phys. Lett. B}
  {\bfseries 69} (1977) 105--108}.

\bibitem{Shao:2012iz}
H.-S. Shao, ``{HELAC-Onia: An automatic matrix element generator for heavy
  quarkonium physics},''
  \href{http://dx.doi.org/10.1016/j.cpc.2013.05.023}{{\em Comput. Phys.
  Commun.} {\bfseries 184} (2013) 2562--2570},
  \href{http://arxiv.org/abs/1212.5293}{{\ttfamily arXiv:1212.5293 [hep-ph]}}.

\bibitem{Shao:2015vga}
H.-S. Shao, ``{HELAC-Onia 2.0: an upgraded matrix-element and event generator
  for heavy quarkonium physics},''
  \href{http://dx.doi.org/10.1016/j.cpc.2015.09.011}{{\em Comput. Phys.
  Commun.} {\bfseries 198} (2016) 238--259},
  \href{http://arxiv.org/abs/1507.03435}{{\ttfamily arXiv:1507.03435
  [hep-ph]}}.

\bibitem{Barbieri:1975ki}
R.~Barbieri, R.~Gatto, R.~Kogerler, and Z.~Kunszt, ``{Meson hyperfine
  splittings and leptonic decays},''
  \href{http://dx.doi.org/10.1016/0370-2693(75)90267-1}{{\em Phys. Lett. B}
  {\bfseries 57} (1975) 455--459}.

\bibitem{Czarnecki:1997vz}
A.~Czarnecki and K.~Melnikov, ``{Two loop QCD corrections to the heavy quark
  pair production cross-section in e+ e- annihilation near the threshold},''
  \href{http://dx.doi.org/10.1103/PhysRevLett.80.2531}{{\em Phys. Rev. Lett.}
  {\bfseries 80} (1998) 2531--2534},
  \href{http://arxiv.org/abs/hep-ph/9712222}{{\ttfamily arXiv:hep-ph/9712222}}.

\bibitem{Beneke:1997jm}
M.~Beneke, A.~Signer, and V.~A. Smirnov, ``{Two loop correction to the leptonic
  decay of quarkonium},''
  \href{http://dx.doi.org/10.1103/PhysRevLett.80.2535}{{\em Phys. Rev. Lett.}
  {\bfseries 80} (1998) 2535--2538},
  \href{http://arxiv.org/abs/hep-ph/9712302}{{\ttfamily arXiv:hep-ph/9712302}}.

\bibitem{Marquard:2014pea}
P.~Marquard, J.~H. Piclum, D.~Seidel, and M.~Steinhauser, ``{Three-loop
  matching of the vector current},''
  \href{http://dx.doi.org/10.1103/PhysRevD.89.034027}{{\em Phys. Rev. D}
  {\bfseries 89} no.~3, (2014) 034027},
  \href{http://arxiv.org/abs/1401.3004}{{\ttfamily arXiv:1401.3004 [hep-ph]}}.

\bibitem{CTEQ:1993hwr}
{\bfseries CTEQ} Collaboration, R.~Brock {\em et~al.}, ``{Handbook of
  perturbative QCD: Version 1.0},''
  \href{http://dx.doi.org/10.1103/RevModPhys.67.157}{{\em Rev. Mod. Phys.}
  {\bfseries 67} (1995) 157--248}.

\bibitem{Wang:2004du}
J.-X. Wang, ``{Progress in FDC project},''
  \href{http://dx.doi.org/10.1016/j.nima.2004.07.094}{{\em Nucl. Instrum. Meth.
  A} {\bfseries 534} (2004) 241--245},
  \href{http://arxiv.org/abs/hep-ph/0407058}{{\ttfamily arXiv:hep-ph/0407058}}.

\bibitem{Butenschoen:2009zy}
M.~Butenschoen and B.~A. Kniehl, ``{Complete next-to-leading-order corrections
  to J/psi photoproduction in nonrelativistic quantum chromodynamics},''
  \href{http://dx.doi.org/10.1103/PhysRevLett.104.072001}{{\em Phys. Rev.
  Lett.} {\bfseries 104} (2010) 072001},
\href{http://arxiv.org/abs/0909.2798}{{\ttfamily arXiv:0909.2798 [hep-ph]}}.
%%CITATION = ARXIV:0909.2798;%%.

\bibitem{Hou:2019efy}
T.-J. Hou {\em et~al.}, ``{New CTEQ global analysis of quantum chromodynamics
  with high-precision data from the LHC},''
  \href{http://dx.doi.org/10.1103/PhysRevD.103.014013}{{\em Phys. Rev. D}
  {\bfseries 103} no.~1, (2021) 014013},
  \href{http://arxiv.org/abs/1912.10053}{{\ttfamily arXiv:1912.10053
  [hep-ph]}}.

\bibitem{ParticleDataGroup:2020ssz}
{\bfseries Particle Data Group} Collaboration, P.~A. Zyla {\em et~al.},
  ``{Review of Particle Physics},''
  \href{http://dx.doi.org/10.1093/ptep/ptaa104}{{\em PTEP} {\bfseries 2020}
  no.~8, (2020) 083C01}.

\bibitem{Denby:1983az}
B.~H. Denby {\em et~al.}, ``{Inelastic and Elastic Photoproduction of J/$\psi$
  (3097)},'' \href{http://dx.doi.org/10.1103/PhysRevLett.52.795}{{\em Phys.
  Rev. Lett.} {\bfseries 52} (1984) 795--798}.

\bibitem{NA14:1986mdd}
{\bfseries NA14} Collaboration, R.~Barate {\em et~al.}, ``{Measurement of
  $J/\psi$ and $\psi^\prime$ Real Photoproduction on $^{6}$Li at a Mean Energy
  of 90-{GeV}},'' \href{http://dx.doi.org/10.1007/BF01548261}{{\em Z. Phys. C}
  {\bfseries 33} (1987) 505}.

\bibitem{Bailey:2020ooq}
S.~Bailey, T.~Cridge, L.~A. Harland-Lang, A.~D. Martin, and R.~S. Thorne,
  ``{Parton distributions from LHC, HERA, Tevatron and fixed target data:
  MSHT20 PDFs},'' \href{http://dx.doi.org/10.1140/epjc/s10052-021-09057-0}{{\em
  Eur. Phys. J. C} {\bfseries 81} no.~4, (2021) 341},
  \href{http://arxiv.org/abs/2012.04684}{{\ttfamily arXiv:2012.04684
  [hep-ph]}}.

\bibitem{NNPDF:2017mvq}
{\bfseries NNPDF} Collaboration, R.~D. Ball {\em et~al.}, ``{Parton
  distributions from high-precision collider data},''
  \href{http://dx.doi.org/10.1140/epjc/s10052-017-5199-5}{{\em Eur. Phys. J. C}
  {\bfseries 77} no.~10, (2017) 663},
  \href{http://arxiv.org/abs/1706.00428}{{\ttfamily arXiv:1706.00428
  [hep-ph]}}.

\bibitem{Altarelli:1988qr}
G.~Altarelli, M.~Diemoz, G.~Martinelli, and P.~Nason, ``{Total Cross-Sections
  for Heavy Flavor Production in Hadronic Collisions and QCD},''
  \href{http://dx.doi.org/10.1016/0550-3213(88)90126-5}{{\em Nucl. Phys. B}
  {\bfseries 308} (1988) 724--752}.

\bibitem{Stevenson:1981vj}
P.~M. Stevenson, ``{Optimized Perturbation Theory},''
  \href{http://dx.doi.org/10.1103/PhysRevD.23.2916}{{\em Phys. Rev. D}
  {\bfseries 23} (1981) 2916}.

\bibitem{Brodsky:1982gc}
S.~J. Brodsky, G.~P. Lepage, and P.~B. Mackenzie, ``{On the Elimination of
  Scale Ambiguities in Perturbative Quantum Chromodynamics},''
  \href{http://dx.doi.org/10.1103/PhysRevD.28.228}{{\em Phys. Rev. D}
  {\bfseries 28} (1983) 228}.

\bibitem{Mojaza:2012mf}
M.~Mojaza, S.~J. Brodsky, and X.-G. Wu, ``{Systematic All-Orders Method to
  Eliminate Renormalization-Scale and Scheme Ambiguities in Perturbative
  QCD},'' \href{http://dx.doi.org/10.1103/PhysRevLett.110.192001}{{\em Phys.
  Rev. Lett.} {\bfseries 110} (2013) 192001},
  \href{http://arxiv.org/abs/1212.0049}{{\ttfamily arXiv:1212.0049 [hep-ph]}}.

\bibitem{Wu:2013ei}
X.-G. Wu, S.~J. Brodsky, and M.~Mojaza, ``{The Renormalization Scale-Setting
  Problem in QCD},'' \href{http://dx.doi.org/10.1016/j.ppnp.2013.06.001}{{\em
  Prog. Part. Nucl. Phys.} {\bfseries 72} (2013) 44--98},
  \href{http://arxiv.org/abs/1302.0599}{{\ttfamily arXiv:1302.0599 [hep-ph]}}.

\bibitem{Grunberg:1980ja}
G.~Grunberg, ``{Renormalization Group Improved Perturbative QCD},''
  \href{http://dx.doi.org/10.1016/0370-2693(80)90402-5}{{\em Phys. Lett. B}
  {\bfseries 95} (1980) 70}. [Erratum: Phys.Lett.B 110, 501 (1982)].

\bibitem{Grunberg:1982fw}
G.~Grunberg, ``{Renormalization Scheme Independent QCD and QED: The Method of
  Effective Charges},'' \href{http://dx.doi.org/10.1103/PhysRevD.29.2315}{{\em
  Phys. Rev. D} {\bfseries 29} (1984) 2315--2338}.

\bibitem{xFitterDevelopersTeam:2018hym}
{\bfseries xFitter Developers' Team} Collaboration, H.~Abdolmaleki {\em
  et~al.}, ``{Impact of low-$x$ resummation on QCD analysis of HERA data},''
  \href{http://dx.doi.org/10.1140/epjc/s10052-018-6090-8}{{\em Eur. Phys. J. C}
  {\bfseries 78} no.~8, (2018) 621},
  \href{http://arxiv.org/abs/1802.00064}{{\ttfamily arXiv:1802.00064
  [hep-ph]}}.

\bibitem{Buckley:2014ana}
A.~Buckley, J.~Ferrando, S.~Lloyd, K.~Nordstr\"om, B.~Page, M.~R\"ufenacht,
  M.~Sch\"onherr, and G.~Watt, ``{LHAPDF6: parton density access in the LHC
  precision era},''
  \href{http://dx.doi.org/10.1140/epjc/s10052-015-3318-8}{{\em Eur. Phys. J. C}
  {\bfseries 75} (2015) 132}, \href{http://arxiv.org/abs/1412.7420}{{\ttfamily
  arXiv:1412.7420 [hep-ph]}}.

\bibitem{Smith:1991pw}
J.~Smith and W.~L. van Neerven, ``{QCD corrections to heavy flavor
  photoproduction and electroproduction},''
  \href{http://dx.doi.org/10.1016/0550-3213(92)90476-R}{{\em Nucl. Phys. B}
  {\bfseries 374} (1992) 36--82}.

\bibitem{Ozcelik:2019qze}
M.~A. Ozcelik, ``{Constraining gluon PDFs with quarkonium production},''
  \href{http://dx.doi.org/10.22323/1.352.0159}{{\em PoS} {\bfseries DIS2019}
  (2019) 159}, \href{http://arxiv.org/abs/1907.01400}{{\ttfamily
  arXiv:1907.01400 [hep-ph]}}.

\bibitem{Mangano:1996kg}
M.~L. Mangano and A.~Petrelli, ``{NLO quarkonium production in hadronic
  collisions},'' \href{http://dx.doi.org/10.1142/S0217751X97002048}{{\em Int.
  J. Mod. Phys. A} {\bfseries 12} (1997) 3887--3897},
  \href{http://arxiv.org/abs/hep-ph/9610364}{{\ttfamily arXiv:hep-ph/9610364}}.

\bibitem{Feng:2015cba}
Y.~Feng, J.-P. Lansberg, and J.-X. Wang, ``{Energy dependence of
  direct-quarkonium production in $pp$ collisions from fixed-target to LHC
  energies: complete one-loop analysis},''
  \href{http://dx.doi.org/10.1140/epjc/s10052-015-3527-1}{{\em Eur. Phys. J. C}
  {\bfseries 75} no.~7, (2015) 313},
  \href{http://arxiv.org/abs/1504.00317}{{\ttfamily arXiv:1504.00317
  [hep-ph]}}.

\bibitem{Lansberg:2021vie}
J.-P. Lansberg, M.~Nefedov, and M.~A. Ozcelik, ``{Matching
  next-to-leading-order and high-energy-resummed calculations of
  heavy-quarkonium-hadroproduction cross sections},''
  \href{http://dx.doi.org/10.1007/JHEP05(2022)083}{{\em JHEP} {\bfseries 05}
  (2022) 083}, \href{http://arxiv.org/abs/2112.06789}{{\ttfamily
  arXiv:2112.06789 [hep-ph]}}.

\end{thebibliography}\endgroup

\providecommand{\href}[2]{#2}\begingroup\raggedright\endgroup

\end{document}